%% file: ms.tex
\documentclass[12pt,preprint]{aastex}

\newcommand{\msun} {$M_{\sun}$}

\newcommand{\Te} {T_{\rm eff}}
\newcommand{\logg} {\log g} 
\def\bra#1{\left\langle #1\right|}
\def\ket#1{\left| #1\right\rangle}

\begin{document}

\title{SPECTROSCOPIC ANALYSIS OF DA WHITE DWARFS:\\
STARK BROADENING OF HYDROGEN LINES\\INCLUDING NON-IDEAL EFFECTS}

\author{P.-E. Tremblay and P. Bergeron}
\affil{D\'epartement de Physique, Universit\'e de Montr\'eal, C.P.~6128, 
Succ.~Centre-Ville, Montr\'eal, Qu\'ebec H3C 3J7, Canada.}
\email{tremblay@astro.umontreal.ca, bergeron@astro.umontreal.ca}

\begin{abstract}

We present improved calculations for the Stark broadening of hydrogen
lines in dense plasmas typical of white dwarf atmospheres. Our new
model is based on the unified theory of Stark broadening from Vidal,
Cooper, \& Smith. For the first time, we account for the non-ideal
effects in a consistent way directly inside the line profile
calculations. The Hummer \& Mihalas theory is used to describe the
non-ideal effects due to perturbations on the absorber from protons
and electrons. We use a truncation of the electric microfield
distribution in the quasi-static proton broadening to take into
account the fact that high electric microfields dissociate the upper
state of a transition. This approach represents a significant
improvement over previous calculations that relied on the use of an
ad hoc parameter to mimic these non-ideal effects. We obtain the
first model spectra with line profiles that are consistent with the
equation of state. We revisit the
properties of DA stars in the range 40,000 K $> \Te >$ 13,000
K by analyzing the optical spectra with our improved models.
The updated atmospheric parameters are shown to differ
substantially from those published in previous studies, with a mean
mass shifted by $+0.034$ \msun. We also show that these revised
atmospheric parameters yield absolute visual magnitudes that 
remain in excellent agreement with trigonometric parallax measurements.

\end{abstract}


\section{INTRODUCTION}

The most successful technique used to determine the effective
temperatures and surface gravities of hydrogen-line DA white dwarfs is
to compare the observed and predicted hydrogen line profiles. This
so-called spectroscopic technique was first applied to a large sample
of DA stars by \citet[][hereafter BSL92]{bergeron92} in the case of
the hydrogen Balmer lines. In recent years, the method has also been
applied to the study of the Lyman line profiles in the ultraviolet
\citep{barstow03,vennes05}. The success of this approach resides in
the fact that the theoretical line profiles are very sensitive to
variations of the atmospheric parameters \citep{wegner81}. This is
illustrated in Figure \ref{fg:f1} where the theoretical profiles of
five Balmer lines are displayed for various values of $\Te$ and
$\logg$. In comparison with the other fitting methods used before (see
BSL92 for an extensive review), the spectroscopic method has the
lowest intrinsic uncertainties, which allows for a more precise
comparison, at least in a relative sense, of the atmospheric
parameters between different DA stars. For instance, BSL92 used the
spectroscopic method to determine the shape of the DA mass
distribution (see also \citealt{liebert05,kepler07}), while
\citet{bergeron95} applied the same method to define the location of
the ZZ Ceti instability strip in a $\logg - \Te$ diagram (see also
\citealt{gianninas06}). Since about 80\% of the white dwarf population
is of the DA type, the spectroscopic technique coupled with high
signal-to-noise spectroscopic observations of the Balmer lines for
large samples of DA stars \citep{gianninas05,sdss} can reveal
important details about the luminosity function and the evolution and
the formation rate of DA white dwarfs \citep{liebert05}.

Even though the {\it relative} accuracy of the spectroscopic method is
mostly limited by the quality of the observations (signal-to-noise
ratio, flux calibration, etc.), the {\it absolute} values of the
atmospheric parameters depend highly on the level of sophistication of
the physics included in the calculations of model atmospheres with
hydrogen-rich compositions. One starting point would be the efficient
code of \citet{wesemael80} that allowed for the computation of
radiative LTE atmospheres for hot DA white dwarfs. Similar codes have
also been developed by D.~Koester and described in
\citet{finley97}. Since then, the most important advance has certainly
been the inclusion of NLTE effects --- in the code of \citet{hubeny95}
among others --- which allowed the study of hot DA stars
($\Te>40,000$~K). In terms of cooler white dwarfs, the code developed
by \citet{bergeron91} made great advances and has been a reference
ever since. It includes convective energy transport and the non-ideal
equation of state of \citet[][hereafter HM88]{hm88} to describe the
atomic populations. Otherwise, it is mostly in the atomic physics,
with new or improved opacities and partition functions, that advances
have been made.

The dominant features observed in DA spectra are the hydrogen
lines. It is thus important to have a good understanding of the atomic
transitions to exploit the power of the spectroscopic technique to its
fullest. The main source of line broadening in most DA stars ($\Te$
$>$ 10,000 K) is due to charged particle interactions, also called
Stark broadening. The theory that has been the most successful to
describe these line profiles is the unified theory of Stark broadening
from Vidal, Cooper, \& Smith (1970, hereafter VCS; see also
\citealt{vcs0,vcs2,vcs3}). This theory was used right from the
beginning in the analysis of large samples of DA white dwarfs (BSL92)
and it is the basis of most of the widely accepted results on the
global properties of these stars. Even though the theoretical
framework of the unified theory has been known since the original work
of VCS, it is not until the calculations of \citet{lemke97} that a
complete grid of hydrogen line profiles from the unified theory became
available, covering the full range of possible transitions,
temperatures, and electronic densities encountered in DA white dwarf
atmospheres. Before that, line profiles from the unified theory were
either extrapolated to high densities (T. Sch\"oning and K. Butler,
private communication), or more drastically, less accurate broadening
theories \citep{underhill59,edmonds67} were used for the highest lines
of the Balmer series (H$\epsilon$ and above). At $\Te$ below $\sim
10,000$~K for the Balmer lines and below 30,000 K for the Lyman lines,
other types of line broadening, mostly due to neutral particles, must
be included in the model atmospheres. The modeling of the
quasi-molecular line opacity \citep{allard04} has been one of the most
significant advance in this domain.

Despite the development of these theoretical tools, the study of DA
stars has suffered from several complications. For instance, in their
preliminary analysis of 129 DA white dwarfs, BSL92 discovered a lack
of internal consistency between the spectroscopic solutions obtained
when an increasing number of Balmer lines were included in the fitting
procedure. This is illustrated at the top of Figure
\ref{fg:f2} for a typical DA white dwarf where we can see that the
solution drifts in the $\Te$ vs $\logg$ diagram as more lines
are included in the fit. This correlation indicates that the physics
included in the model calculations needed significant
improvement. \citet{bergeron93} traced back the problem to the neglect
of non-ideal effects inside the Stark broadening
calculations. Indeed, \citet{seaton90} had argued that non-ideal
effects, such as those taken into account in the HM88 equation of
state, have also to be included directly in the line profile
calculations to get a coherent physical framework. However, it was not
possible at the time of the analysis of BSL92 to rework
the line broadening calculations. Instead, the authors chose to
include an ad hoc parameter inside the model atmosphere code to {\it
mimic} the non-ideal effects in the line profiles
\citep{bergeron93}. By taking twice the value of the critical electric
microfield ($\beta_{\rm crit}$, see \S~2.3) from the HM88 theory, it was
found that the internal consistency improved substantially, as can be
seen in the middle panel of Figure \ref{fg:f2}. {\it It should be
stressed, however, that this does not imply that HM88 underestimated
the value of the critical field}. This is just a quick and dirty way
to simulate the non-ideal effects by reducing the line wing opacity,
in particular where the line wings overlap (see \citealt{bergeron93}
for further details). This ad hoc parameter has been used ever
since in all model spectra of white dwarf stars
\citep[BSL92;][]{hubeny95,finley97,vennes06}.

In this paper, we solve the problem discussed above in a more elegant
way, by computing improved Stark broadening profiles based of the
unified theory of VCS, but by including non-ideal effects directly in
the line profile calculations following an approach similar to that
originally proposed by \citet{seaton90}. The non-ideal effects due to
the proton and electron perturbations are taken into account using the
HM88 theory. We begin in \S~2 by discussing the two successful
theories at the heart of our calculations, namely the VCS unified
theory of Stark broadening and the HM88 non-ideal equation of
state. In \S~3, we describe how these two formalisms are combined for
the first time to compute Stark broadening profiles that take into
account non-ideal effects. We then evaluate in \S~4 the implications
of our improved line profiles on the spectroscopic analysis of DA
stars. Our conclusions follow in
\S~5.

\section{LINE BROADENING THEORY}

The basis of model atmospheres for DA white dwarfs resides in the
equation of state and opacity calculations of the hydrogen gas. One of
the most successful equations of state used to compute the populations
of the different accessible states (bound states of hydrogen, H$^-$,
molecular H$_2$, H$_2^{+}$ and H$_3^{+}$, as well as proton and
electron populations) is the HM88 non-ideal equation of state
described in more detail in section \S~2.2. With these populations in
hand for all states of the gas, we can proceed to calculate the gas
opacity, which is generally split into true absorption and scattering
processes. In particular, the bound-bound opacity (or line opacity)
between one initial state $i$ and one final state $j$ is written as

\begin{equation}\label{e2}
\kappa_{ij} (\nu)\,d\nu = N_{i}\frac{\pi e^2}{m_{e}c}f_{ij} \phi(\nu)\,d\nu ~,
\end{equation}

{\noindent}where $N_{i}$ is the population of the initial state and
$f_{ij}$ is the oscillator strength of the transition. These two factors
define the amplitude of the transition while $\phi(\nu)$ is the
spectral broadening profile normalized to unity. In this paper, we are
mostly interested in the broadening profile for the hydrogen
lines. This opacity is the key ingredient of the spectroscopic method
that relies on a detailed comparison of observed and predicted line
profiles. 

Generally speaking, the line profile is a convolution of different
broadening mechanisms. The first source of broadening is due to the
interaction between the absorber and the charged particles in the
plasma.  This process is referred to as Stark broadening and is
discussed in more detail in section
\S~2.1. Stark broadening represents the dominant source of broadening
for most transitions in DA white dwarfs. Accounting only for this
source of broadening, the line profile can be expressed
\citep{underhill59,edmonds67} as

\begin{equation}\label{e3}
\phi (\nu)\,d\nu = \frac{\lambda^2}{cF_{0}}S^*(\alpha)\,d\nu ~,
\end{equation}

{\noindent}where $\alpha = \Delta \lambda / F_{0}$ with
$\Delta \lambda$ measured from the center of the
line. $F_{0}$ is defined as the electric field at the mean
distance between the plasma ions

\begin{equation}\label{e4}
F_0 = Z_{p} e (4\pi N_{p}/3)^{2/3} = 1.25 \times 10^{-9} Z_{p} N_{e}^{2/3} ~,
\end{equation}

{\noindent}where $Z_p$ and $N_p$ are the charge and density of
perturbers, respectively. The profile $S^*(\alpha)$ is 
a convolution of a pure Stark profile $S(\alpha)$ and a Voigt function
$H(a,v)$ that takes into account thermal and natural
broadening

\begin{equation}\label{e5}
S^*(\alpha)=\displaystyle\int^{\infty}_{-\infty}S\left(\alpha+\frac{\lambda_{0}v}{F_{0}c}\right)\frac{H(a,v)}{\sqrt{\pi}}dv ~,
\end{equation}

{\noindent}where $a=\Gamma/4\pi\Delta\nu_{D}$, with $\Gamma$
representing the natural broadening half-width and $\Delta\nu_{D}$ the
Doppler width. The second type of broadening important in DA white
dwarfs is the interaction of the absorber with neutral hydrogen. It is
especially important at low effective temperatures for transitions
involving an upper level with a low principal quantum number. In the
case of Balmer lines, we must include resonance broadening
\citep{ali65,ali66} and nonresonant broadening \citep{hammond91} in
cool white dwarfs with $\Te$ $<$ 10,000 K. Since this source of
broadening has a Lorentzian profile, the broadening parameter can be
added directly to that of the natural broadening. For the lower Lyman
lines (L$\alpha$, L$\beta$, and L$\gamma$), close range collisions of
the absorber with hydrogen atoms and protons cause the appearance of
important satellites in the wings of the lines that are visible up to
$\Te \sim 30,000$ K
\citep{allard82,allard04}. This opacity also affects the thermodynamic
structure of the atmosphere. Furthermore, at very low effective
temperatures, H-H$_{2}$ collisions become the main source of
broadening for L$\alpha$ \citep{kowalski06}.

\subsection{Stark Broadening}

The Stark effect is defined as the shifting --- or splitting --- of
spectral lines under the action of an electric field. In the following
section, we assume that we have in the atmospheric plasma a local
electric microfield due to protons that is constant with time. This
static approximation is well justified since the characteristic time
for the fluctuation of the local microfield due to a change in the
proton distribution is much larger than the characteristic time of
absorption and emission processes that we are studying
\citep{stehle93}. We also assume that the microfield is constant in
space, a good approximation if the perturbers are far from the area
occupied by the bound electron orbits. The regime where this
approximation fails is discussed further in \S~2.3. The electric
microfield, assumed to be along the $z$-axis, interacts with the
dipolar moment of the atom and leads to a perturbation in the
Hamiltonian of the form
 
\begin{equation}\label{e6}
H_{s}=e\vec{R} \cdot \vec{F}=eFz ~.
\end{equation}

It is common to write the amplitude of the microfield in terms of the
unitless parameter $\beta$ such that
\begin{equation}\label{e7}
\beta=\frac{F}{F_0} ~,
\end{equation}

{\noindent}where $F_0$ is the characteristic field defined by equation
(\ref{e4}). The amplitude of the microfield can be expressed using a
probability distribution $P(\beta)$. One such distribution that can be
computed analytically is the Holtsmark distribution, which takes into
account the interactions (vectorial sum) between the different
perturbers. We use here the more physical distribution of
\citet{hooper68} that takes into account the Debye screening effect,
and which corresponds also to the formulation included in the work of
VCS.  HM88 originally used the Holtsmark distribution, but for
consistency we use the approach of \citet{nayfonov99} for the HM88
equation of state by replacing the Holtsmark distribution with the
Hooper distribution. We note that this improved version of the HM88
equation of state is already included in our model atmosphere code,
although it has never been properly documented.

We note that the perturbation $H_{s}$ does not commute with the
angular momentum $\vec{L}^2$ since it has a privileged direction in
space. The effect of breaking this symmetry is to lift the degeneracy
in energy of the hydrogen atomic levels. There are no exact solutions
to this problem and we generally use quantum mechanical perturbation
theory to determine corrections to the energy and wave functions. The
first order corrections to the energy are linear in relation to the
electric microfield that is applied, and represent the basis of the
linear Stark effect. The linear corrections are found by diagonalizing
the perturbation operator in the degenerate subspaces. For a hydrogen
atom, we can write this expression analytically. First of all, knowing
that the operator $z$ conserves the azimuthal quantum number $m$, we
can diagonalize the matrix $\bra{\Psi_{k}}H_{s}\ket{\Psi_{l}}$ 
\citep{condon} by making use of the quantum numbers $(n,m,q)$, where
$q=q_{1}-q_{2}$ with $q_{1}+q_{2}=n-|m|-1$, and $n$ is the principal
quantum number. The shifts in energy are then given by
 
 \begin{equation}\label{e9}
\Delta E^{1}_{n,m,q}=\frac{3ea_{0}F}{2}nq ~.
\end{equation}

When the electric microfield becomes very high, or equivalently, when
the Stark splitting is important enough that the Stark components of
two levels with a different principal quantum number are crossing, the
linear approximation is no longer valid. The second order corrections
to the energy become quadratic in relation to the microfield, and we
find

\begin{equation}\label{e10}
\Delta E^{2}_{n,m,q}=e^2F^2\sum_{n'\not=n}\frac{|\bra{\Psi_{n',m',q'}}z\ket{\Psi_{n,m,q}}|^{2}}{E^{0}_{n,m,q}-E^{0}_{n',m',q'}} ~.
\end{equation}

The problem with this perturbation approach to the Stark effect is
that the wave functions, using any order of the perturbation theory,
are not necessarily normalizable \citep{friedrich06}. This is because
the perturbing potential goes to $-\infty$ when $z$ goes to
$-\infty$. In other words, states that were bound without a microfield
are now only metastable states, and there is a finite probability that
the atom will be ionized when the microfield has a sufficiently high
amplitude. Classically, the sum of the electric potentials of the
absorber and the nearby protons only allows bound states in the local
potential minima up to a certain energy, called the saddle point. Part
of this problem is because we have used a spatially uniform microfield
(eq. \ref{e6}). In reality, the microfield cannot be uniform beyond a
distance of about the interparticle separation, and thus under normal
conditions, most of the states will be bound. However, for sufficiently
high microfields, some excited states will become unbound and this is where
resides the problem discussed further in the next sections.
 
\subsection{The Unified Theory of Stark Broadening}

In addition to static protons, the absorber also interacts
with free electrons. These particles being much faster than the
protons, a collisional approach is generally used to describe this
interaction. Since the collisions are rapid, the interaction is
non-adiabatic and a proper quantum mechanical treatment of the
collisions accounting for the internal structure of the hydrogen atom
is required. Various theories of Stark broadening for hydrogen lines
take into account both the electron and proton interactions
\citep{vcs2,seaton90,stehle93}. Here we summarize the principal aspects
of the unified theory of VCS, which even today stands as the most
accurate theory for the conditions encountered in white dwarf
atmospheres.

The unified theory of VCS uses the quasi-static proton broadening
approximation with the distribution of electric microfields described
in \S~2.1, assumed to be constant in space and time. The
complete Stark profile is then defined as

\begin{equation}\label{e11}
S(\alpha)=\displaystyle\int^{\infty}_{0}P(\beta)I(\alpha,\beta)d\beta ~,
\end{equation}

{\noindent}where $I(\alpha,\beta)$ is the electronic broadening
profile. Therefore, the problem is reduced to the calculation of
electronic broadening profiles for a fixed microfield of amplitude
$\beta$. The full broadening profile is then the average of the
electronic profiles over all possible microfields weighted by the
probability distribution $P(\beta)$. The unified theory is a
non-adiabatic quantum mechanical theory that uses the classical path
approximation for the electron-absorber interactions, implying that
the wave functions are well separable. The interaction potential
considers only the first dipole term. The name of the unified theory
comes from the fact that the electronic profile is valid at all
detunings from the line center. In the asymptotic limit of the line
core, the impact approximation is recovered. Furthermore, in the far
wings of the lines, the one-electron theory is recovered and the
electrons and protons have similar contributions to the broadening.

Now we want to compute the electronic profile for an initial Stark
state $n_a$ to a final state $n'_a$, where this abbreviated
notation is used to designate any of the $(n,m,q)$ states.  Mostly to
reduce the computing time and to simplify the inversion of the matrix
involved (see below), the unified theory uses the no-quenching
approximation, which means that there are only collision-induced
transitions within the same principal level. The states after the
collisions will be denoted by $n_b$ and $n'_b$ for the initial and
final state, respectively. The calculations can then be performed by
considering only two-level transitions. This last approximation is
valid only if the difference in energy between the highest Stark state
from level $n'$ and the lowest state of level $n'+1$ is large, or in
other words, when the different lines are well separated (i.e., when
the line wings do not overlap). Only then is the regime of the linear
Stark effect consistent with the basic assumptions of the unified
theory.

The electronic broadening profile $I(\alpha,\beta)$ is given by the
sum over all states

\begin{equation}\label{e12}
I(\alpha,\beta)=\frac{1}{\pi}\sum_{n_a,n'_a,n_b,n'_b}\rho(n_a)\,{\rm Im}\left[\bra{n_a}\vec{d}\ket{n'_a}\bra{n'_b}\vec{d}\ket{n_b}\bra{n_b}\bra{n'_b}\mathcal{K}^{-1}(\alpha,\beta)\ket{n'_a}\ket{n_a}\right] ~,\end{equation}

{\noindent}where $\vec{d}$ is the dipole operator, and $\rho$ is the
probability that the atom is in the initial state (Boltzmann
factor). The operator $\mathcal{K}$ takes into account the linear
Stark effect as well as the free electron-atom interactions
\citep[see][eq. XII.2 for more details]{vcs1}. Since the operator
$\mathcal{K}$ is generally not diagonal in the basis of the $(n,m,q)$
states, it induces a coupling between the various Stark components.

\subsection{Non-Ideal Equation of State}

One important advance that has provided a better interpretation of the
hydrogen lines in DA white dwarfs is a realistic modeling of non-ideal
effects at high densities. \citet{bergeron91} were the first to
include the occupation probability formalism of HM88 to determine the
populations of the bound states of hydrogen in white dwarf
atmospheres. This probabilistic approach considers the perturbations
on each atom by charged and neutral particles. One advantage of this
statistical interpretation is that there are no discontinuities in the
populations and the opacities when the temperature and the pressure
vary, in contrast with other formalisms that simply predict the last
bound atomic level under given physical conditions.

Briefly, each atomic level $n$ has a probability $w_n$ of being bound and a
probability $1-w_n$ of being dissociated due to perturbations from other
particles in the plasma. The partition function of a given species
is then written as

\begin{equation}\label{e13}
Z=\sum_{n} w_n g_n \exp\left( -\frac{\chi_n}{kT} \right) ~,
\end{equation}

{\noindent}where $\chi_n$ and $g_n$ are respectively the excitation
energy and multiplicity of the level. The occupation probability $w_n$
is defined as

\begin{equation}\label{e14}
w_n=\exp\left( -\frac{\partial f /\partial N_n}{kT} \right) ~,
\end{equation}

{\noindent}where $f$ is the free energy of the non-ideal
interaction. The different interactions being statistically
independent, the total occupation probability can then be calculated
simply as the product of the contributions from neutral and charged
particles.

The interaction with neutral particles is treated within a hard sphere
model. The configurational free energy is derived from the second
virial coefficient in the van de Waals equation of state (excluded
volume correction). The exact origin of this excluded volume term is
the hard sphere equation of state of \citet{carnahan69}. The
interaction with charged particles is the one closely related to the
Stark effect. This interaction suggests that the electric microfields,
fluctuating on long time scales due to changes in the spatial
distribution of protons, can destabilize a bound state and cause its
dissolution into the continuum. Indeed, we already discussed in \S~2.1
that the application of an intense electric field on a bound atom
allow for the ionization of the atom. HM88 therefore suggest an
occupation probability for proton perturbations of the form

\begin{equation}\label{e15}
w_n({\rm charged})=\displaystyle\int^{\beta_{\rm crit}}_{0}P(\beta)\,d\beta ~,
\end{equation}

{\noindent}where $P(\beta)$ is the probability distribution of the
electric microfields introduced in \S~2.1. This occupation probability
implies that all microfields larger than the critical field
$\beta_{\rm crit}$ will ionize the electrons in level $n$ (it is
implicit here that the critical field depends on the atomic level
considered). The difficulty with this formulation is to find the 
value of the critical microfield, for which slightly different
expressions can be found in the literature
\citep[HM88;][]{seaton90,stehle93}. The simplest formulation is that
of \citet{seaton90} which consists in taking the critical microfield
as the point where the energy of the highest Stark state for a given
level $n$ crosses the energy of the lowest Stark state of the next
level $n+1$. For hydrogen, and using the linear Stark effect, we find

\begin{equation}\label{e16}
\beta _{\rm crit}=\frac{2n+1}{6n^4(n+1)^2}\frac{e}{a_0^2 F_{0}} ~.
\end{equation}

The assumption that the crossing of two atomic levels with different
principal quantum numbers leads to the dissolution of the lower level
is difficult to prove and it is based in part on laboratory
experiments and theoretical considerations (see HM88 for a more
detailed discussion). Let us consider one bound electron in one of its
many Stark states of the $n$-th level. When the electric microfield,
which fluctuates in time, gets to a value such that there is a {\it
crossing}\footnote{Strictly speaking, direct degenerate crossings are
avoided due to fine structures and non-uniformity in the microfields.}
between this level and the lowest Stark state of the $n+1$ level,
there is a significant probability that when the microfield goes down,
there will be a transition $n \rightarrow n+1$. The incessant
fluctuations of the microfield imply that these transitions will
continue until the electron becomes unbound. However, the bound
electron will in fact be in a rather homogeneous superposition of all
its accessible Stark states due to electronic collisions\footnote{See
equation \ref{e12}; the non-diagonal operator $\mathcal{K}$ introduces this
coupling.} on time scales much shorter than the fluctuation time for
the microfields, and also due to fluctuations in the direction of the
microfield. That is, as soon as the first Stark state from a level $n$
crosses another one from the level $n+1$, the electron is allowed to
cascade to the continuum.

The reader might have noticed that equation (\ref{e16}) is based on the linear
Stark effect although this is not necessarily a good approximation
when levels are crossing. This is why HM88 considered a formulation of
the critical field that takes into account corrections of higher order
from the perturbation theory as well as results from
experiments\footnote{Remember that even if we have the {\it exact}
critical field, ionization does not necessarily occur exactly at that
point and this is why experimental data are also required.}. They find
that for levels with $n>3$, the linear theory remains valid. For lower
levels, however, the critical field is large enough that it must have
been created by a single proton very close to the absorber. In this
case, the approximation of a uniform electric field in space fails and
we must consider the full potential curve of the H$_{2}^{+}$ system to
find the critical field \citep{stehle93}, and also to compute the
broadening profiles of the far line wings
\citep{allard82}. Exact calculations show that no crossing is possible
between levels $n=1$, 2, and 3, which implies that L$\alpha$ is not
affected by non-ideal effects due to proton perturbations. Also, the
crossing between $n=3$ and 4 is close to the classical saddle point
value, which means that the linear Stark theory is in error by $\sim
14\%$.  Therefore, HM88 added a smoothing factor to equation (\ref{e16}) so
that the exact result is recovered for $n=3$. For levels with a high
value of $n$, HM88 obtain the same results as
\citet{seaton90}. For the Balmer lines, these different values of
$\beta_{\rm crit}$ are not an issue since the non-ideal effects
discussed above are never important for H$ \alpha$ in DA white
dwarfs. This is not the case for L$\alpha$ and L$\beta$, however, for
which non-ideal effects should be treated with caution, or simply
neglected (for instance in TLUSTY and in our code discussed in \S~4).

We have seen in \S~2.2 that the interaction of hydrogen atoms with
rapid electrons was better described with a collisional approach since
the quasi-static theory is generally not valid in this case. These
perturbations are also considered in the HM88 model. We must keep in
mind that because the plasma is in thermal equilibrium, the equation
of state already accounts for the inelastic electronic collisions that
cause transitions and ionization. However, we expect that if the time
between collisions is small, the energy of a bound level will be left
undefined by a certain amount given by the uncertainty principle. If
this energy uncertainty is of the order of the ionization potential,
we expect the state to become unbound after the collision. As stated
in HM88, this contribution to the occupation probability is always
much lower than the proton contributions and can be neglected in DA
atmospheres. However, the electronic perturbations should not be
neglected when calculating the line profiles, as discussed in \S~3. We
must also remember that the electrons contribute indirectly to the
occupation probability due to proton perturbations by mixing the
Stark states. The occupation probability for the bound states of
hydrogen are displayed in Figure \ref{fg:f3} for typical conditions
encountered at the photosphere of white dwarf stars. Non-ideal effects
are shown to be extremely important except for the very lowest
transitions. The solution of the HM88 equation of state yields the
populations for all states.

Finally, the opacity calculation must also be modified to take into
account the occupation probability formalism, as described by
\citet{dappen87}. The bound-bound (line) opacity for a transition
between levels $i$ and $j$ (equation \ref{e2}) is multiplied by
$w_{j}/w_{i}$ to take into account the fact that these levels
can be dissolved when a photon is absorbed. We thus obtain

\begin{equation}\label{e17}
\kappa_{ij} (\nu)\,d\nu = N_{i}\frac{\pi e^2}{m_{e}}\frac{w_j}{w_i} f_{ij} \phi(\nu)\,d\nu ~.
\end{equation}

{\noindent}When the absorption is from a bound state to a final state
that has been dissolved by non-ideal interactions (with a probability
$1-w_{j}/w_{i}$), we obtain instead a bound-free opacity since this
process is equivalent to the ionization of the atom. We can directly
extrapolate, in this case, the bound-free cross section at frequencies
below the usual cutoff frequency $\nu_{c}$. 
\citet{dappen87} have treated this opacity by
considering the absorption from a level $i$ of a photon of energy $h\nu$
that yields a transition to a fictitious upper level $n^\ast$ given by

\begin{equation}\label{e18}
n^\ast=\left(\frac{1}{n_{i}^2}-\frac{h\nu}{\chi^I}\right)^{-1/2} ~.
\end{equation}

{\noindent}The hydrogen bound-free opacity for $\nu < \nu_{c}$ --- also
called the pseudo-continuum opacity --- can then be written as

\begin{equation}\label{e19}
\kappa_{i} (\nu) d\nu = N_{i}\left(1-\frac{w_{n^\ast}}{w_i}\right) \frac{64 \pi^4 m_{e} e^{10}}{3 \sqrt{3} c h^6} \frac{g_{bf,i}(\nu)}{n_{i}^5 \nu ^3}\,d\nu ~.
\end{equation}

{\noindent}In Figure \ref{fg:f4}, we show the individual contributions
of the line and pseudo-continuum opacities
at the photosphere of a 20,000 K DA white dwarf in the spectral region
of the Balmer lines. We can see how the pseudo-continuum opacity
extends to wavelengths longward of the Balmer jump, potentially
affecting the opacity between the lines, and in particular the high
Balmer lines.

\section{STARK BROADENING PROFILES INCLUDING NON-IDEAL EFFECTS}

The VCS line profiles (\S~2.2) and the HM88 equation of state (\S~2.3)
have long been included in modern white dwarf model atmospheres
\citep[BSL92;][]{hubeny95,finley97,vennes06}. The HM88
theory has been used, in particular, to calculate populations as well
as line and pseudo-continuum opacities. However, all model spectra
currently available and calculated within the HM88 framework suffer
from a major inconsistency. \citet{seaton90} was the first to point
out that not only the line strengths will be affected by non-ideal
effects, but the {\it shape} of the line profiles as well. The Stark
broadening profiles $S(\alpha)$ from VCS, for instance, do not take
into account such non-ideal effects. It is possible to describe
qualitatively the behavior of non-ideal line profiles. The high
electric microfields, normally responsible for the absorption in the
far wings of the line profiles, will not contribute as much to the
line opacity in the non-ideal case since these fields are mostly
responsible for dissolving the atomic levels. Consequently, the
opacity in the line wings will be significantly reduced with respect
to the ideal case. But since by definition the profiles are normalized
to unity, the non-ideal line profiles will have deeper cores and
appear narrower in comparison with the ideal case.

Instead of including the non-ideal effects discussed above,
\citet{bergeron93} proposed a solution that could be easily
implemented in model atmosphere calculations. As mentioned in the
Introduction, this is the solution adopted by BSL92 in their analysis
of a large spectroscopic sample of DA stars. The authors found in
their preliminary analysis that the atmospheric parameters varied
significantly when an increasing number of Balmer lines was included
in the fitting procedure. Furthermore, the mean mass of their sample
was uncomfortably low \citep[$\left<M\right>\sim
0.53$~\msun;][]{bergeron90a}. The solution proposed by
\citet{bergeron93} was to parameterize the value of the critical field
in equation (\ref{e16}). The effect of varying the value of $\beta_{\rm crit}$ is,
among other things, to change the relative importance of the line wing
and pseudo-continuum opacities. By increasing {\it arbitrarily} the value of
the critical field, one mimics the non-ideal effects by reducing the
line wing opacity. One problem with this approach, of course, is that
the determination of the multiplicative factor in the expression for
$\beta_{\rm crit}$ has absolutely no physical basis. This factor can only
be determined from an empirical analysis of the internal consistency
of the fitting procedure, similar to that shown in Figure {\ref{fg:f2}
\citep[see also][]{bergeron93}. Furthermore, there is no reason to 
expect this ad hoc parameter to be the same for all hydrogen levels
(i.e., Balmer or Lyman series) or for different atmospheric
parameters. Another side effect of this approach is to artificially
change the atomic populations in the HM88 equation of state, which in
turn could have an unexpected impact on other parts of the model
calculations. It should be clear by now that a more physical approach
is seriously required, especially given the high quality spectra that have
become available in recent years, either in the visible or the
ultraviolet. This ad hoc parameter represents an important hurdle
towards our understanding of the global properties of white dwarfs.

A coherent way to combine the Stark broadening profiles and the HM88
equation of state is described at length in \citet{seaton90}. Seaton
proposed to replace equation (\ref{e11}) by

\begin{equation}\label{e20}
S(\alpha)=\frac{\displaystyle\int^{\beta_{\rm crit}}_{0}P(\beta)I(\alpha,\beta)d\beta}{\displaystyle\int^{\beta_{\rm crit}}_{0}P(\beta)d\beta} ~,
\end{equation}

{\noindent}where only electric microfields $\beta$ with an amplitude
inferior to the critical field $\beta_{\rm crit}$ for the upper level
of the transition now contribute to the broadening of the lines by the
protons. The higher fields contribute instead to the pseudo-continuum
opacity. The denominator, which turns out to be the occupation
probability due to proton perturbations, allows for the
renormalization of the line profiles\footnote{In the definition of
\citet{seaton90}, the line profiles are not renormalized but instead
the factor $w_j$ is omitted in equation \ref{e17}.}.

\citet{seaton90} was the first to calculate Stark broadening profiles
taking into account non-ideal effects. However, his calculations made
for the Opacity Project rely on an approximate electron broadening
theory that is inappropriate in the context of white dwarf
atmospheres. \citet{stehle93} also discuss the implementation of
non-ideal effects inside the line profile calculations using an
alternative electron broadening theory. This theory also lifts in part
the static approximation for the protons and therefore gives better
results in the center of the profiles, but this has no effect on the
spectrum at the high densities found in white dwarf atmospheres. In
the line wings, the results are equivalent to the VCS formulation. We
must mention, however, that all published tables and subsequent papers
for this theory
\citep{stehle94,stehle99} do not include non-ideal effects.

In this work, we use the Seaton approach (equation \ref{e20}) to include
non-ideal effects due to proton perturbations inside the unified
theory of Stark broadening from VCS. We also include non-ideal
corrections due to electronic perturbations according to the HM88
theory. This was neglected by \citet{seaton90}, likely because it was
estimated that the effects would be negligible in the context studied. As
discussed in \S~2.3, the collisions between free electrons and the
absorber leave a bound level in a state of indefinite
energy. Therefore, the HM88 model states that if the uncertainty on
the energy becomes equal to the ionization energy of the level, the
electron has a good probability of becoming unbound. Here we have at
hand the electronic broadening profiles given by equation (\ref{e12}), and
their wings generally extend to infinity in terms of the detuning
$\Delta \nu$ (measured from the line center). However, according to
the HM88 theory, it is obvious that the detuning induced by the
electronic collisions cannot extend much beyond $\nu_c$ (i.e.~the
ionization threshold frequency for this level) or there would be line
opacity shortward of the Balmer or Lyman jumps. An abrupt cutoff is
not desirable since this could cause discontinuities in the
spectra. We adopt here an exponential cutoff so that the electronic
broadening profiles are reduced by a factor of $e^{-1}$ at $\Delta
\nu=\nu_c-\nu_0$, where $\nu_0$ is the central frequency of the
transition, and then we renormalize the total broadening
profiles. Since the exponential factor is independent of $\beta$, we
can easily represent this correction in the form

\begin{equation}\label{e20b}
\phi^{\prime}(\nu)=\frac{\phi(\nu)e^{-|\nu-\nu_0|/\nu_c}}{\displaystyle\int^{\infty}_{0}
\phi(\nu)e^{-|\nu-\nu_0|/\nu_c}\,d\nu}\ .
\end{equation}

\noindent
This has a small but non-negligible effect on the line wings,
especially for the higher members of the series. Both $\Te$ and
$\logg$ are increased by $\sim 0.5$ \% when this correction is
introduced in our calculations. Using alternative types of cutoff, we
estimate that the uncertainties on this part of the theory can be as
high as half of the shifts found here. As mentioned in HM88, a
more physical approach is certainly much needed but such a theory still does
not exist.

Figure \ref{fg:f5} presents the Stark broadening profiles
$S(\alpha)$ for the Balmer lines H$\gamma$ and H8 at typical
photospheric conditions of DA stars. Along
with the results from this work, we show the original calculations of
VCS \citep{lemke97} and the approximate profiles from
\citet{seaton90}. We note that our profiles including non-ideal
effects due to proton and electron perturbations are very different
from the original ideal gas calculations of VCS, even though we are
using the same broadening theory. Our profiles have significantly less
opacity in the line wings, and higher lines of the series have much
sharper profiles, as anticipated. To have a better idea of the
relative importance of the two improvements discussed above, we have
estimated that the electronic corrections account for 12\% and 16\% of
the reduction of the opacity at 100 \AA\ from the line center of
H$\gamma$ and H8, respectively. Finally, we show that our profiles are
not in agreement with the calculations of
\citet{seaton90}, which include non-ideal effects due to proton
perturbations but use only an approximate treatment for the electronic
broadening. These results confirm that this last approximation is not
suitable for typical conditions encountered in DA white dwarf
atmospheres. The comparison of our Lyman line profiles with
the calculations of VCS or Seaton is similar to those
displayed in Figure \ref{fg:f5}.

We have demonstrated in the previous sections that our approach to
Stark broadening provides significant physical improvements over
previously available calculations. When included in a model
atmosphere, the equation of state and the opacity calculations become
entirely consistent without the need of the ad hoc parameter introduced
by \citet{bergeron93}. By including non-ideal effects, we have shown
that the line profiles from the unified theory are dramatically
modified in typical white dwarf atmospheres. These improved profiles
now need to be validated using spectroscopic data.

\subsection{Comparison with Laboratory Experiments}

The preferred way to validate Stark broadening theories for hydrogen
has been the comparison with laboratory plasma emissivity measurements
from \citet{wiese72}. These pure hydrogen plasma arc experiments were
performed at high densities and temperatures, comparable to the
photospheric conditions of cool white dwarfs. The main advantage of
this experiment is that it is not under the constraint of radiative
equilibrium, unlike a stellar atmosphere. In other words, the emergent
flux at one wavelength is not affected by what occurs at other
wavelengths. The emissivity can then be simply written as

\begin{equation}\label{e21}
j_{\lambda}(\lambda)=\frac{2 c}{\lambda^4}\kappa(\lambda)e^{-hc/\lambda kT} ~,
\end{equation}

{\noindent}where $\kappa(\lambda)$ represents the total monochromatic opacity. Another
advantage, at least from a theoretical point of view, is that it is
easier to control the experiments and the results are independent of
various sources of uncertainty intrinsic to DA model atmospheres
(i.e., convection, contamination from heavy elements, etc.). We use
the data for the lowest and highest density experiments (the most
extreme positions on the plasma arc) from Wiese et al., which were
compared to broadening theories in various studies
\citep{dappen87,seaton90,bergeron93,stehle93}. Wiese et al.~estimate
the LTE plasma parameters for these two experiments at $\log T=4.00$,
$\log N_{e}=16.26$ and $\log T=4.12$, $\log N_{e}=16.97$,
respectively. However, there is no simple way to measure the plasma
state parameters and it is not clear whether LTE is reached or
not. The authors used the total intensity from two lines, together
with the intensity at two continuum points, and fitted the data with
an approximate plasma model. Since then, slightly different parameters
have been used to compare with broadening theories. Instead of
choosing approximate values for the parameters like in previous
analyses, we performed a $\chi^2$ fit to the full data sets, using the
same input physics as for our white dwarf models. The results with the
original VCS calculations and our improved line profiles are displayed
in Figures \ref{fg:f6} and
\ref{fg:f7} in terms of absolute predicted fluxes (i.e., without any
renormalization).

For the lower density experiment of Figure \ref{fg:f6}, our improved
profiles provide a much better fit to the laboratory data, in
particular in the regions of the high Balmer lines where the wings
overlap. The higher density experiment of Figure \ref{fg:f7} is more
problematic. We find that there is a partial $T-N_e$ degeneracy in the
$\chi^2$ diagram; it is indeed possible to find many acceptable
solutions from visual inspection by increasing both $T$ and
$N_e$. However, to provide the best overall fit with our new profiles,
we have to increase the plasma state parameters significantly compared
to the Wiese et al.~values. We see that our profiles provide the best
fit for the higher lines, although the red wings of H$\beta$ and
H$\gamma$ are predicted a bit too weak. All in all, we conclude that
the experiments of Wiese et al.~do not provide such a stringent
constraint on the broadening theories because of the large range of
acceptable plasma state parameters, and also because of potential
departures from LTE.

\subsection{Further Theoretical Improvements}

While our work represents a significant advance over previous
calculations, second order effects in the theory of Stark broadening
could still come into play. Such second order effects have been
studied extensively in the literature in the case of high density
hydrogen plasma experiments \citep{lee98,stehle00,demura08}. The
sources of these second order effects are plentiful but they all have
the same impact on the line profiles: they produce asymmetries and
they shift the central wavelengths, the latter being unimportant for
the spectroscopic technique used to measure the atmospheric parameters
of white dwarfs. The recent study of \citet{demura08} reveals that the
various second order effects compete and interact together in a
complex way and that they should all be included in the models
simultaneously. The most important effects are due to quadrupole
proton-absorber interactions and to the quadratic Stark effect, but
second order effects due to quadrupole electron-absorber interactions
are noticeable as well. The lifting of the no-quenching approximation
could also be considered a second order effect, although it was shown
by \citet{lee98} to have an impact mostly on the line cores.
Furthermore, some of these quenching effects have already been
accounted for implicitly by including here the HM88 non-ideal effects
directly inside the line profile calculations. Finally, even if we
include better physics for the atomic transitions, we still have to
rely on the HM88 equation of state, which is, or may become, the main
source of uncertainty.

Our analysis above of the Wiese et al.~experiments, with our improved
profiles has already revealed the existence of line profile
asymmetries. For the lower density experiment (see Fig.~\ref{fg:f6}),
it is only a mild effect in the far wings, however. It is not clear if
such second order effects would be significant in a typical DA white
dwarf at $\logg\sim8$, for which the electronic density at the
photosphere is roughly halfway between the two Wiese et
al.~experiments. Therefore, it is probably premature at this stage to
include second order effects in our calculations, and it is not clear
whether such changes would affect our line profiles significantly. We
come back to this point in the analysis of the PG spectroscopic
sample discussed below.

\section{APPLICATION TO WHITE DWARF ATMOSPHERES}

\subsection{Model Spectra}

We now discuss the astrophysical implications of our improved line
profiles described in \S~3 on the modeling of DA white dwarfs. Our
model atmosphere code is based on the program originally developed by
\citet{bergeron91,bsw95} and references therein. The main difference
is that we have updated several sources of opacity and partition
functions (all within the occupation probability formalism). For
completeness, we provide in Table 1 the complete list of opacity
sources included in our code. We have not included the L$\alpha$ line
broadening due to H$_2$ collisions
\citep{kowalski06} since these calculations are still not available, 
although this has no impact over the range of effective temperature
considered here. 

We restrain our analysis to effective temperatures above $\sim
13,000$~K to avoid additional uncertainties related to convective
energy transport. Indeed, the atmospheric structure of DA white dwarfs
below this temperature depends sensitively on the assumed convective
efficiency. The most commonly used approach to include convection in
model atmosphere calculations, the mixing-length theory, is at best a
very crude approximation, and the convective efficiency must be
parameterized by carefully adjusting the value of the mixing length
\citep{bergeron95}. It is also believed that large amounts of helium
can be brought to the surface by convection while remaining
spectroscopically invisible \citep{bergeron90b,tremblay08}. Hence, it
is not even clear whether cool DA stars have hydrogen-rich
atmospheres.  Another reason for restricting our analysis to hotter
stars is that neutral line broadening becomes important for the Balmer
lines below $\Te \sim 10,000$~K, preventing us from performing a
direct comparison of the observed and predicted Stark broadened
absorption lines. We also set an upper limit of $\Te = 40,000$~K
because NLTE effects become important above this temperature. Also,
absorption from heavier elements, particularly in the UV, is likely to
affect the atmospheric structure of hot white dwarfs. 

We thus computed two grids of model atmospheres, one with the VCS profiles
and the other with our improved profiles (both calculated with
$\beta_{\rm crit} \times 1$), covering the range of $T_{\rm
eff}=12,000$~K to 45,000 K (with steps of 500 K up to 15,000 K, 1000 K
up to 18,000 K, 2000 K up to 30,000 K, and 5000 K above) and of
$\logg=6.5$ to 9.5 (by steps of 0.5 dex with additional models at 7.75
and 8.25). To compare the model spectra calculated from both sets of
line profiles, we simply fit the VCS spectra with our improved
spectra. The results of this exercise are illustrated in Figure
\ref{fg:f8}. In general, our new models yield systematically higher
effective temperatures (by $1000-2000$~K) and surface gravities (by $\sim
0.1$~dex) over the entire range of atmospheric parameters considered
here. One exception is a region at low temperatures and high surface
gravities where the trend in temperature is reversed. The strip,
inclined in the diagram near $\Te\sim 15,000$~K, where this separation
occurs corresponds to the region where the strength of the Balmer
lines reaches its maximum
\citep{bergeron95}. Our new models predict Balmer lines that are
stronger, hence the temperatures in this region are pushed towards
lower values.

\subsection{Analysis of the DA Stars in the PG Sample}

In this section, we measure the implications of our improved Stark
profiles on the analysis of Balmer line observations of DA stars by
considering the Palomar-Green sample of
\citet{liebert05}. This sample of 348 DA
stars has been studied in great detail, and the range of effective
temperatures for these objects corresponds very well to that
considered in our analysis. About 250 white dwarfs fall in the
appropriate range of $\Te$, depending on which model grid is used.
The data set and the fitting procedure of the optical spectra are
identical to those described at length in \citet[][and references
therein]{liebert05}. Briefly, we first normalize the observed and
model flux from each line to a continuum set to unity at a fixed
distance from the line center. The observed profiles are then compared
with the predicted profiles, convolved with a Gaussian instrumental
profile. The atmospheric parameters are then obtained using the
non-linear least-squared method of Levenberg-Marquardt, fitting
simultaneously five lines (H$\beta$ to H8). In some cases where the
spectrum is contaminated by a M dwarf companion, one or two lines are
excluded from the fit.

As discussed in the Introduction, model spectra calculated with
standard Stark broadening profiles yield inconsistent atmospheric
parameters when different lines are included in the fitting procedure.
As shown in the bottom panel of Figure \ref{fg:f2}, our improved Stark
profiles provide an even better internal consistency than the previous
calculations displayed in the two upper panels. To better quantify
this internal consistency, we perform for each star in the PG sample
the same exercise as that shown in Figure \ref{fg:f2} using the VCS
profiles and our improved profiles. We then compute for the 250 stars
in our sample the average absolute deviations in $\Te$ and $\logg$
between the solutions obtained with a different number of lines
included in the fit. The results of this exercise are presented as
filled circles in Figure \ref{fg:f9} together with the mean
uncertainties from the fitting procedure as a reference point (the
open circles will be discussed in \S~4.3). We can see that the
deviations with our improved profiles have been significantly reduced
by a factor of $\sim 1.6$ in $\Te$ and $\sim 1.8$ in $\logg$.  Furthermore,
these deviations now both lie within the mean uncertainties of the
fitting procedure, a result which is extremely reassuring.

The quality of the fits is also an important aspect of the comparison
between model grids. We compare in Figure \ref{fg:f10} our best fit to
WD 0205+250 (same star as in Fig.~\ref{fg:f2}) using the VCS profiles
and our improved Stark profiles. This comparison reveals that while
the atmospheric parameters are significantly different, the quality of
the fits is similar. We thus conclude that the quality of the fits
cannot help to discriminate between both sets of Stark profiles.

We can also explore the global properties of the PG sample with our
two model grids. We first convert the $\logg$ values into mass using
evolutionary models appropriate for white dwarfs with thick hydrogen
layers (see \citealt{liebert05} for details). Our results are
presented in Figure \ref{fg:f11} in a mass versus effective
temperature diagram. As expected from the previous comparison
displayed in Figure \ref{fg:f10}, our improved line profiles yield
significantly larger masses and higher effective temperatures. We note
that the results displayed in the bottom panel are qualitatively
similar to the early spectroscopic determination of the mass
distribution of DA stars by \citet{bergeron90a}, which was based on
the VCS profiles without any modification of the critical field. In
this case, the mass distribution has a mean value near 0.53 \msun,
which is uncomfortably low compared to what is expected from earlier
phases of stellar evolution \citep[see the discussion
in][]{bergeron90a}.

Finally, we have also investigated whether second order effects in the
Stark broadening theory, such as those discussed in
\S~3.2, could be detected in our analysis. In particular, if such 
effects are present, we would expect to observe differences in the
fits of the blue and red wings of the Balmer lines.  We have thus
reanalyzed the PG sample by including in the fit (1) only the red wing
of the Balmer lines and (2) only the lines cores (this is accomplished
by fitting only half of the wavelength range we normally use for each
line). We find that the atmospheric parameters obtained in this manner
are entirely consistent, and we thus conclude that second order
effects can be safely neglected in DA white dwarfs.

\subsection{A Reappraisal of Previous Studies of DA White Dwarfs}

The spectroscopic determination of the mass distribution by BSL92 was
based on model spectra that include the solution proposed by
\citet{bergeron93} to mimic the non-ideal effects, namely by taking
twice the value of the critical field ($\beta_{\rm crit} \times 2$) in
the HM88 formalism. This ad hoc procedure had the effect of
artificially reducing the pseudo-continuum opacity, and thus the
opacity in the wings of the high Balmer lines. The internal
consistency of the solutions obtained from different lines was
consequently improved, as can be judged from the results displayed in
Figures \ref{fg:f2} and
\ref{fg:f9} (open circles). Since this ad hoc solution has been adopted by BSL92 and
in all white dwarf models used in the literature, it is important to
evaluate the differences between our revised atmospheric parameters
and those published in the literature based on these approximate model
spectra. For this purpose, we have also calculated another model
atmosphere grid using the original VCS profiles from \citet{lemke97}
but with twice the value of the critical field. This grid is similar
to that used by the Montreal group in the past ten years or so, and
which has been applied to several studies of DA white dwarfs using the
spectroscopic technique.

Once again, we rely on the PG sample of DA stars analyzed above. A
comparison similar to that shown in Figure \ref{fg:f11} is not very
instructive here since the differences in $\Te$ and $\logg$ are
considerably smaller. Instead, we use the representation displayed in
Figure \ref{fg:f12} where the differences $\Delta\Te$ and
$\Delta\logg$ are shown as a function of effective temperature. This
comparison reveals that our improved line profiles yield higher
effective temperatures and surface gravities, with an important
correlation with $\Te$. In particular, the $\logg$ values are about
0.05 dex larger above 20,000~K but can be as much as 0.1 dex larger
near 15,000~K.  Similarly, the differences in temperatures reach a
maximum near 25,000~K but decrease at both lower and higher effective
temperatures. The corresponding mass distributions, displayed in
Figure \ref{fg:f13}, indicate that the mean mass is shifted by
$+0.034$ \msun\ when our new models are used. The shape of the mass
distributions is statistically equivalent, however, and the dispersion
remains the same. We must also point out that according to the results
of \citet{tremblay08}, $15\%$ of the DA white dwarfs in the
temperature range considered here probably have thin hydrogen layers
($M_{\rm H}/M_{\rm tot} < 10^{-8}$). Consequently, the typical values
for the mean mass are probably $\sim 0.005$ to 0.01 \msun\ lower than
the values reported in Figure \ref{fg:f13}.

As discussed in the Introduction, the spectroscopic technique provides
very accurate measurements of the atmospheric parameters, allowing for
a relative comparison of these parameters among individual
stars. However, the {\it absolute} values of the atmospheric
parameters may suffer from an offset due to uncertainties in the
physics included in the model calculations. Hence it is very important
to compare the results with independent methods. The most reliable
independent observational constraint for DA white dwarfs comes from
trigonometric parallax measurements. \citet{holberg08a} have shown
using the recent photometric calibrations of \citet{holberg06} that
there exists a very good correlation between spectroscopically based
photometric distance estimates and those derived from trigonometric
parallaxes. Here we compare absolute visual magnitudes
instead of distances. We first combine trigonometric parallax
measurements with $V$ magnitudes to derive $M_V(\pi)$ values. We then
use the calibration of \citet{holberg06} to obtain $M_V({\rm spec})$
from spectroscopic measurements of $\Te$ and $\logg$. We selected 92
DA stars with known parallaxes from the sample of \citet{bergeron07}
(the uncertainties on the parallaxes must be less than 30\%). The
$M_V$ values obtained from both model grids described in this section
are compared in Figure \ref{fg:f14}. The agreement is very good within
the parallax uncertainties for both model grids. Hence, despite the
fact that our new models yield higher values of $T_{\rm eff}$ and
$\log g$ (i.e., smaller radii), these two effects almost cancel each
other and the predicted luminosities (or $M_V$) remain unchanged, and
so are the conclusions of \citet{holberg08a}. Another important
constraint is provided by the bright white dwarf 40 Eri B for which a
very precise trigonometric parallax and visual magnitude have been
measured by Hipparcos. These measurements yield $M_V=11.01
\pm 0.01$, while we predict $M_V=11.02 \pm 0.07$ and $10.97
\pm 0.07$ based on our new and VCS models, respectively. Although both
determinations agree within the uncertainties with the observed value,
our new grid provides an exact match to the measured $M_V$ value.

We also point out that trigonometric parallax measurements are
available mostly for cool white dwarfs \citep{bergeron97} and nearby
white dwarfs \citep{holberg08b}, which are excluded from our analysis.
In contrast with our approach here, \citet{bergeron97} and
\citet{holberg08b} make use of photometric measurements that cover the
full spectral energy distributions, and both studies find mean masses
near 0.65-0.66 \msun, consistent with our spectroscopic determinations for
the PG sample.

\section{CONCLUSION}

We have combined for the first time in a consistent physical framework
the unified theory of Stark broadening from Vidal, Cooper, \& Smith
and the HM88 non-ideal equation of state. Both of these well
known theories represent the basis of our current knowledge of DA
white dwarfs. Following the suggestion of the late Mike Seaton, we
have taken into account the non-ideal effects due to proton and
electron perturbations directly into the line profile calculations. We
have shown that our improved line profiles are significantly different
from other Stark profiles that have been used in previous studies. We
have computed the first grid of model spectra without the need of the
parameterization of the critical field introduced by
\citet{bergeron93} and used in {\it all} previous models to mimic
these non-ideal effects. We have demonstrated that our new profiles
have important astrophysical implications. In particular, the mean
mass of DA white dwarfs is $\sim 0.03$ \msun\ higher than previously
measured. Yet, our updated atmospheric parameters determined from the
spectroscopic technique remain in excellent agreement with the
constraints imposed by trigonometric parallax measurements.

Future work will confront our improved models with observations in the
more complex regime of higher and lower effective temperatures. An
exhaustive look at the Lyman line analysis of UV observations from FUSE is
also much needed since the model spectra are highly sensitive to the
non-ideal effects in that particular spectral region. For cooler white
dwarfs, we must also investigate the abrupt cut-off that has been
introduced to limit the effects of the Lyman pseudo-continuum opacity.

\acknowledgements
We thank A.~Gianninas for a careful reading of our manuscript, and
M.~Lemke for providing us with his computer version of the VCS
code. We also thank the two referees for their constructive comments,
which have greatly helped improving the presentation of our
results. This work was supported in part by the NSERC Canada and by the
Fund FQRNT (Qu\'ebec). P. Bergeron is a Cottrell Scholar of Research
Corporation for Science Advancement.

\clearpage
\include{tab1}

\clearpage

\figcaption[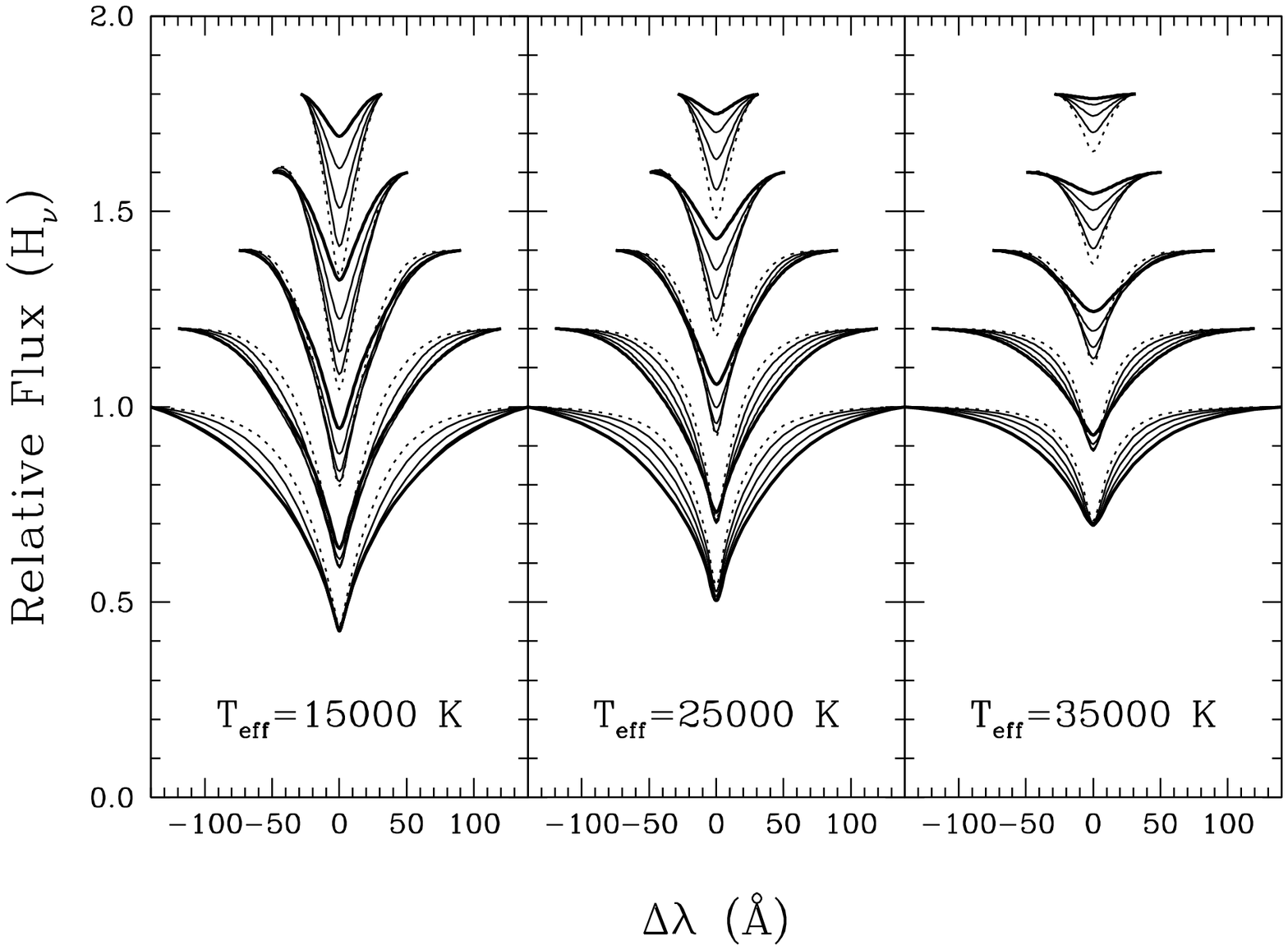] 
{Theoretical line profiles of models at different effective
temperatures and surface gravities using our new line profiles
discussed in \S~3. The lines correspond to H$\beta$ to H$8$ ({\it bottom to
top}) from the Balmer series of the hydrogen atom. In each panel, the
line profiles range from $\logg=7.0$ ({\it dashed line}) to 9.0 ({\it
thick line}) by steps of 0.5 dex. The profiles have been 
convolved with a 6 \AA\ FWHM Gaussian profile, normalized to a continuum
set to unity, and offset vertically from each other for clarity.
\label{fg:f1}}

\figcaption[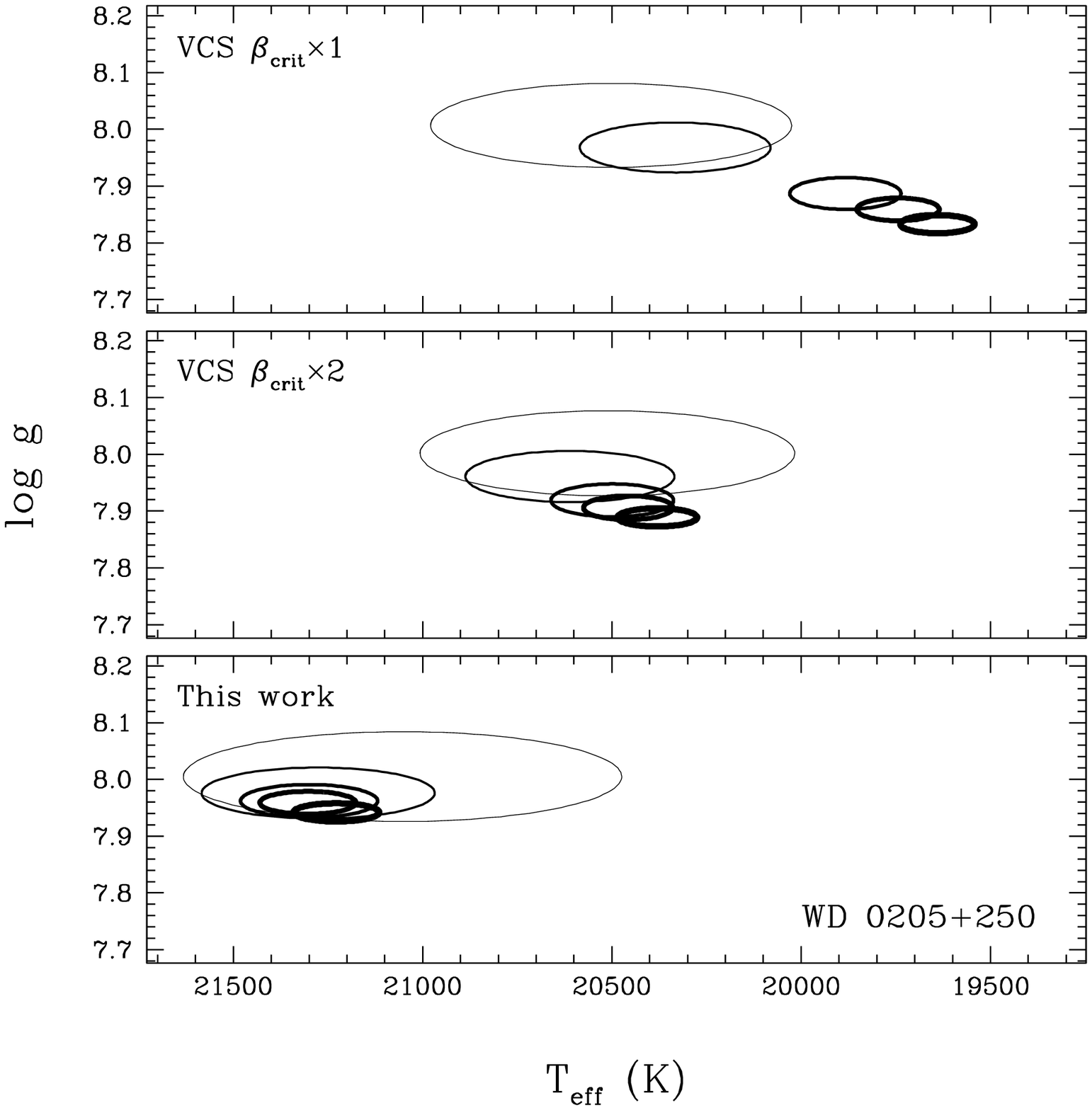] 
{Solutions in a $\Te - \logg$ diagram for a typical DA star using 1
line (H$\beta$), 2, 3, 4 and 5 lines (up to H8) in the fitting
procedure (represented by thicker 1 $\sigma$ uncertainty ellipses from
our fitting procedure). The top panel shows the results with the VCS
line profiles, while the middle panel also includes the ad hoc parameter
proposed by \citet{bergeron93}. The bottom panel is with our new line
profiles discussed in this work (\S~3).
\label{fg:f2}}

\figcaption[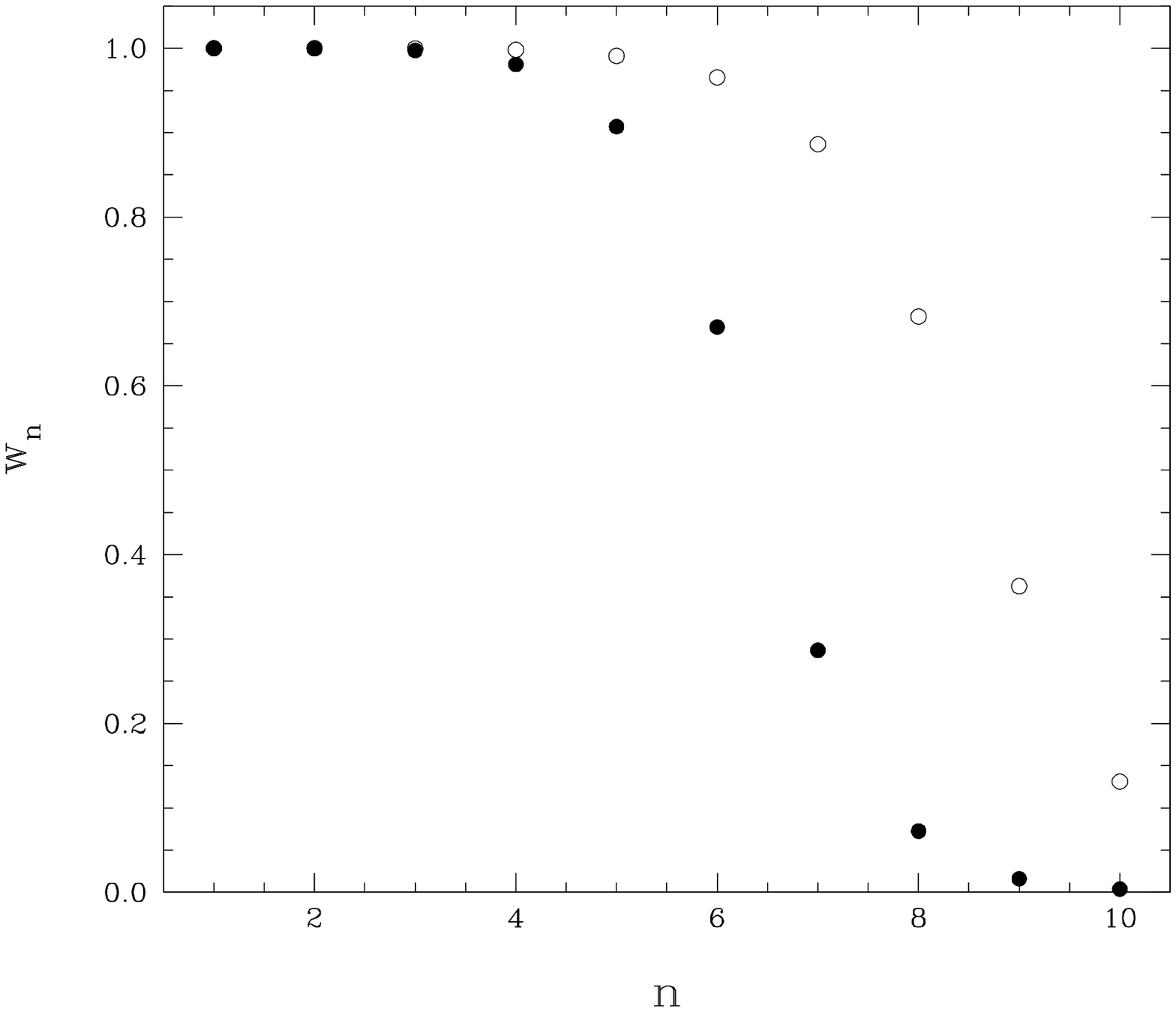] 
{Total occupation probability $w_n$ for the bound states of hydrogen
with principal quantum number $n$ for $T=10,000$~K, $\log N_e=17$
({\it filled circles}) and $T=20,000$~K, $\log N_e=16$ ({\it open circles}).
\label{fg:f3}}

\figcaption[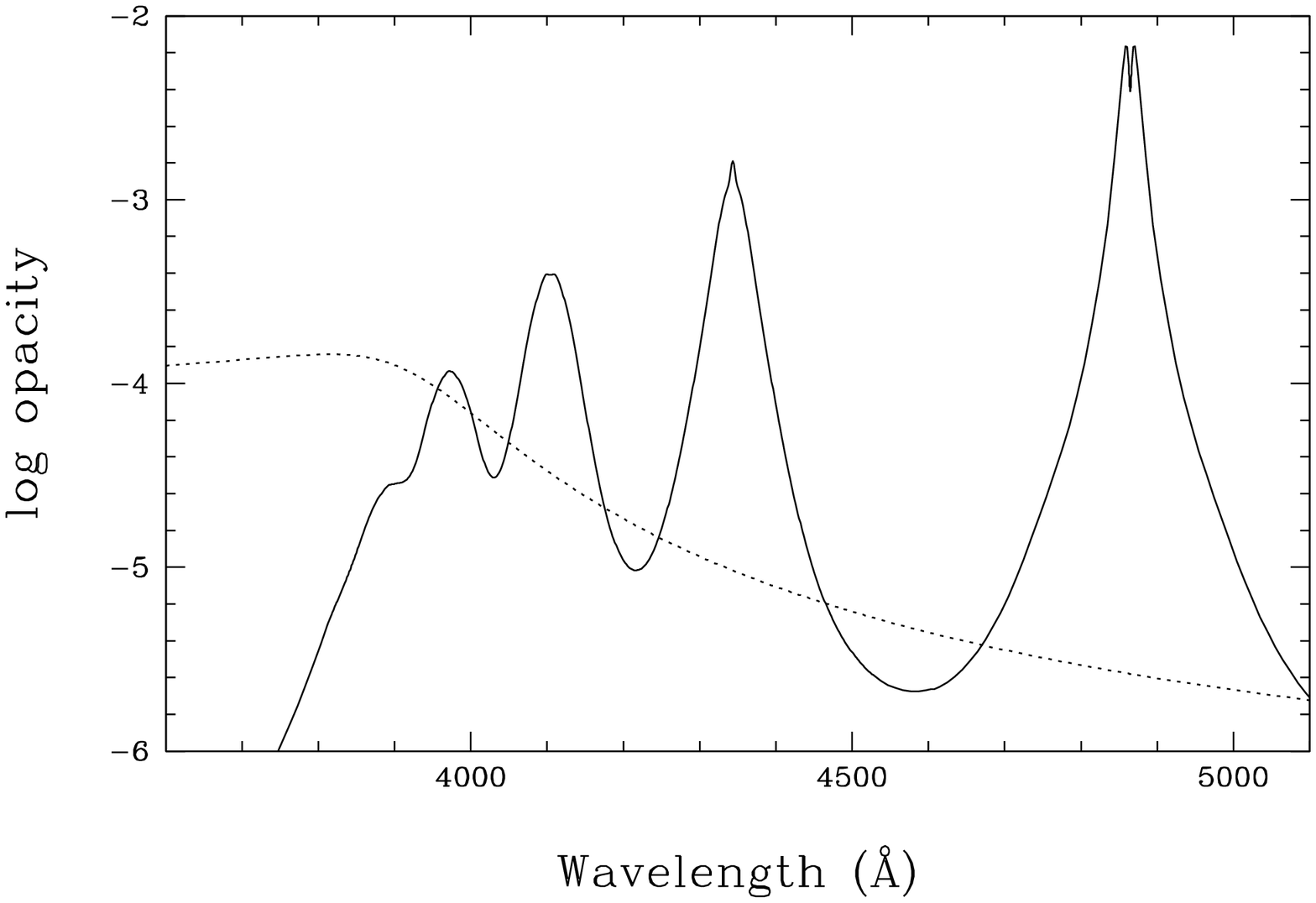] 
{Atomic hydrogen opacity at the photosphere of a 20,000 K DA white
dwarf. The contributions of line and pseudo-continuum opacities are
shown as solid and dotted lines, respectively.
\label{fg:f4}}

\figcaption[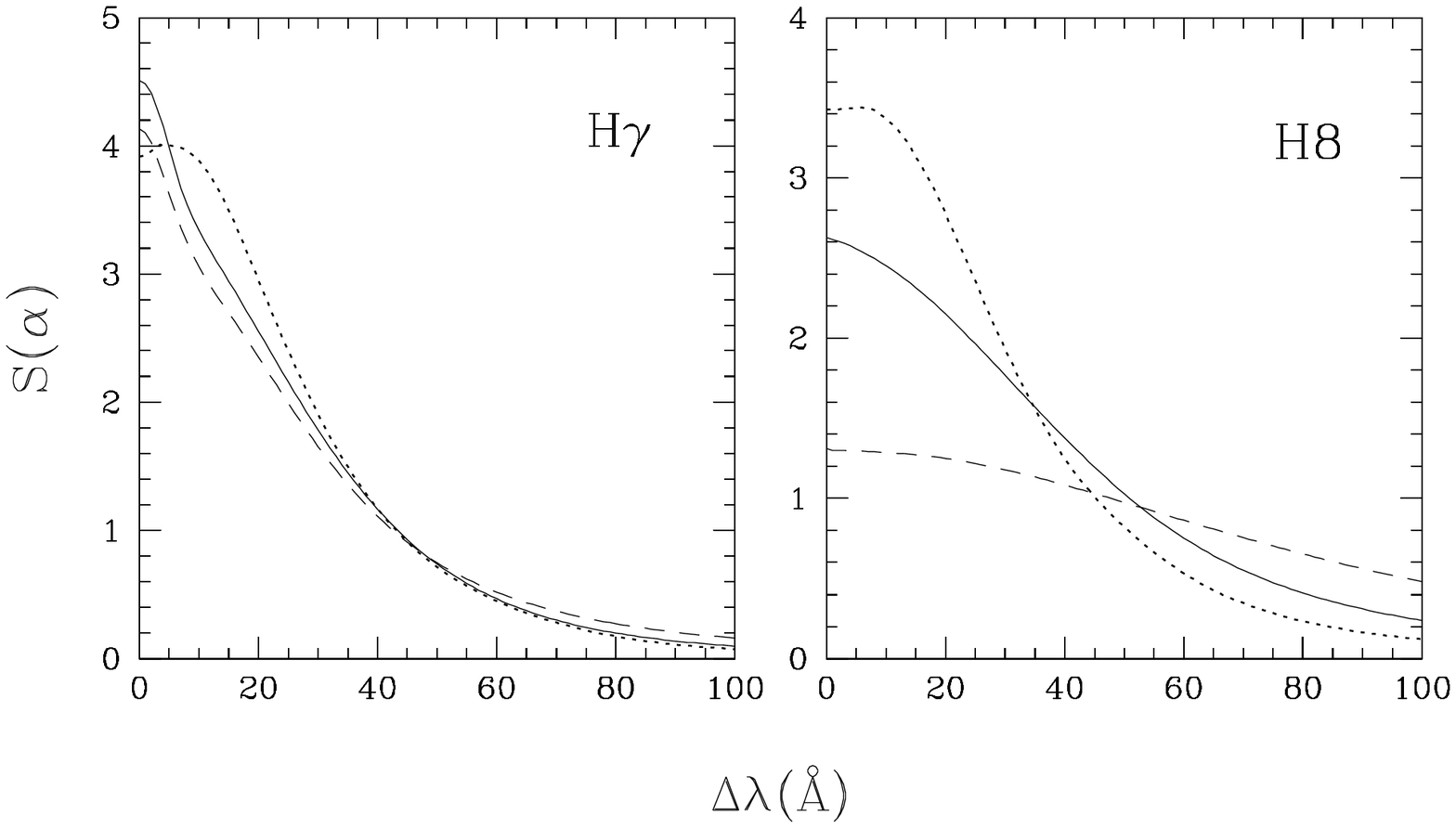] 
{Stark broadening profiles as a function of wavelength measured from
the line center at $T=10,000$~K and $\log N_e=17$. Results are shown
for H$\gamma$ ({\it left}) and H8 ({\it right}). In each panel, we
compare the results from this work (based on the VCS theory coupled
with non-ideal effects from HM88; {\it solid lines}) with the
original VCS calculations ({\it dashed lines}). Also shown are the
approximate calculations of \citet[][{\it dotted lines}]{seaton90},
which also include non-ideal effects.
\label{fg:f5}}

\figcaption[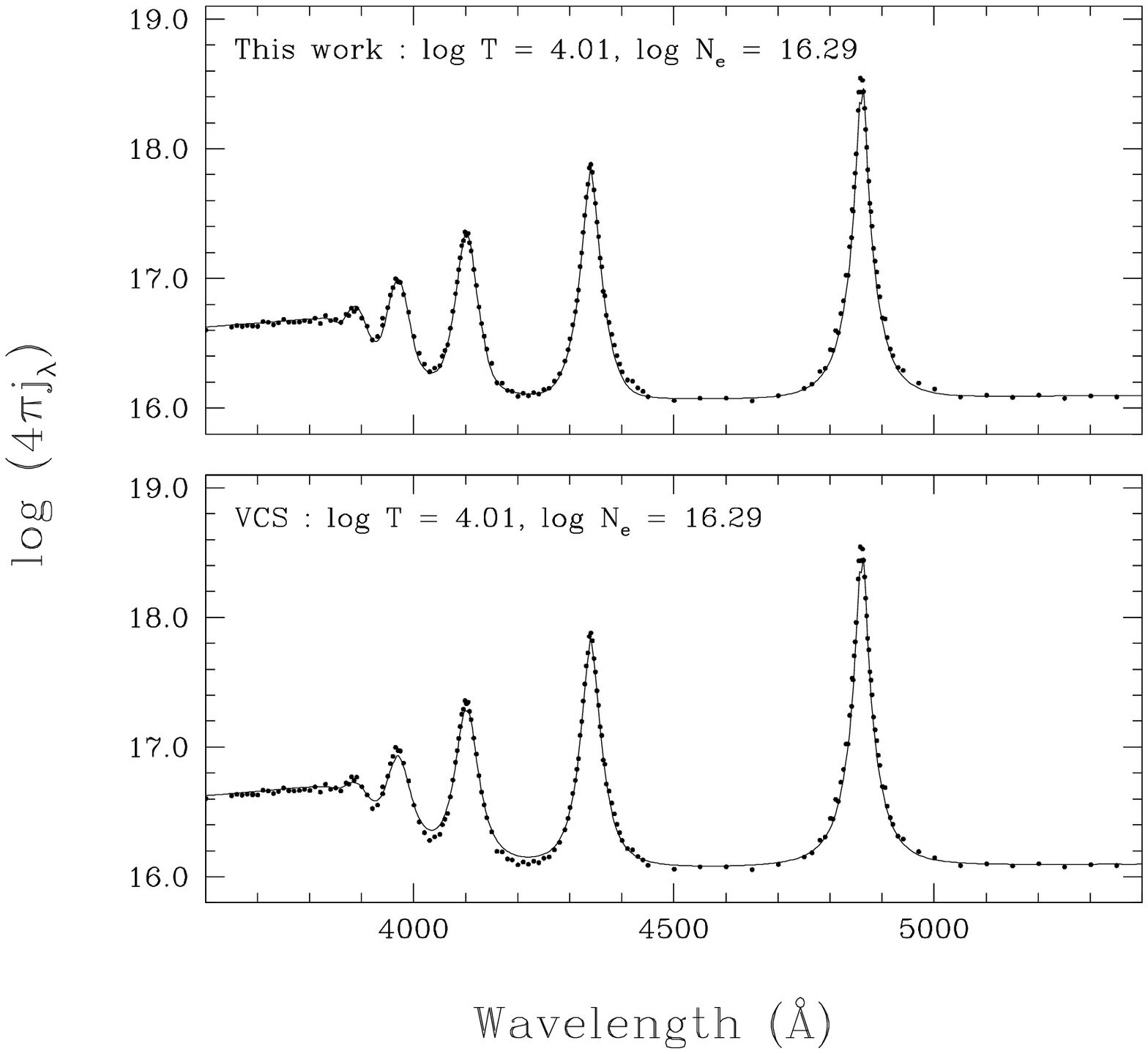] 
{Emissivity of a hydrogen plasma ({\it filled dots}) for the low-density
experiment of \citet{wiese72}. Our best fits ({\it solid line}) to the data are
shown for different line broadening theories identified at the
top of each panel along with the plasma state parameters obtained from
the minimization procedure.
\label{fg:f6}}

\figcaption[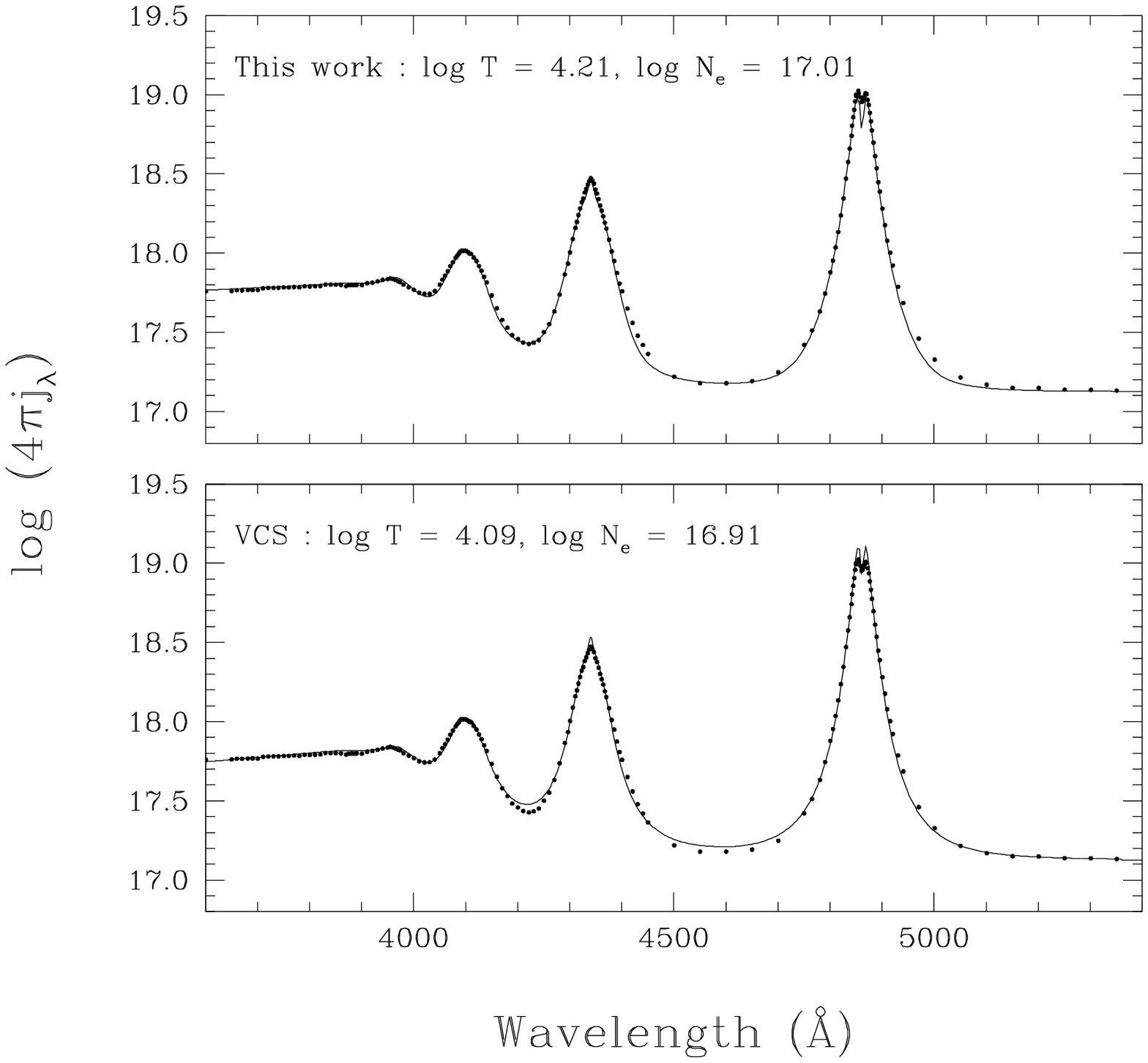] 
{Same as Figure 6 but with the data set for the high-density
experiment of \citet{wiese72}.
\label{fg:f7}}

\figcaption[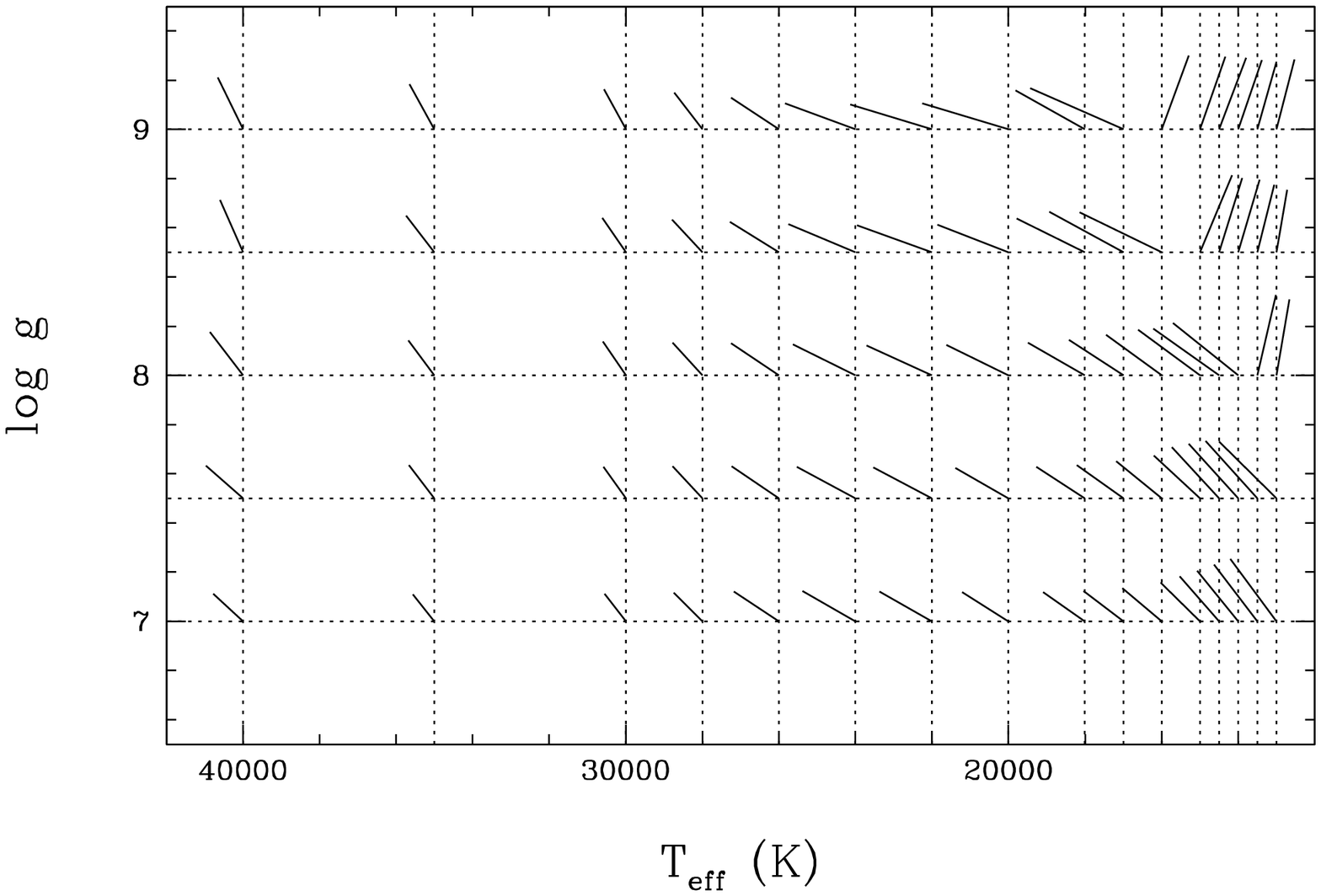] 
{Corrections that must be applied to transform the atmospheric
parameters obtained from the VCS profiles to our improved models. These
have been obtained by simply fitting the Balmer line profiles of the
VCS model grid with our new spectra. For clarity, we have omitted the
grid points at $\logg =7.75$ and 8.25.
\label{fg:f8}}

\figcaption[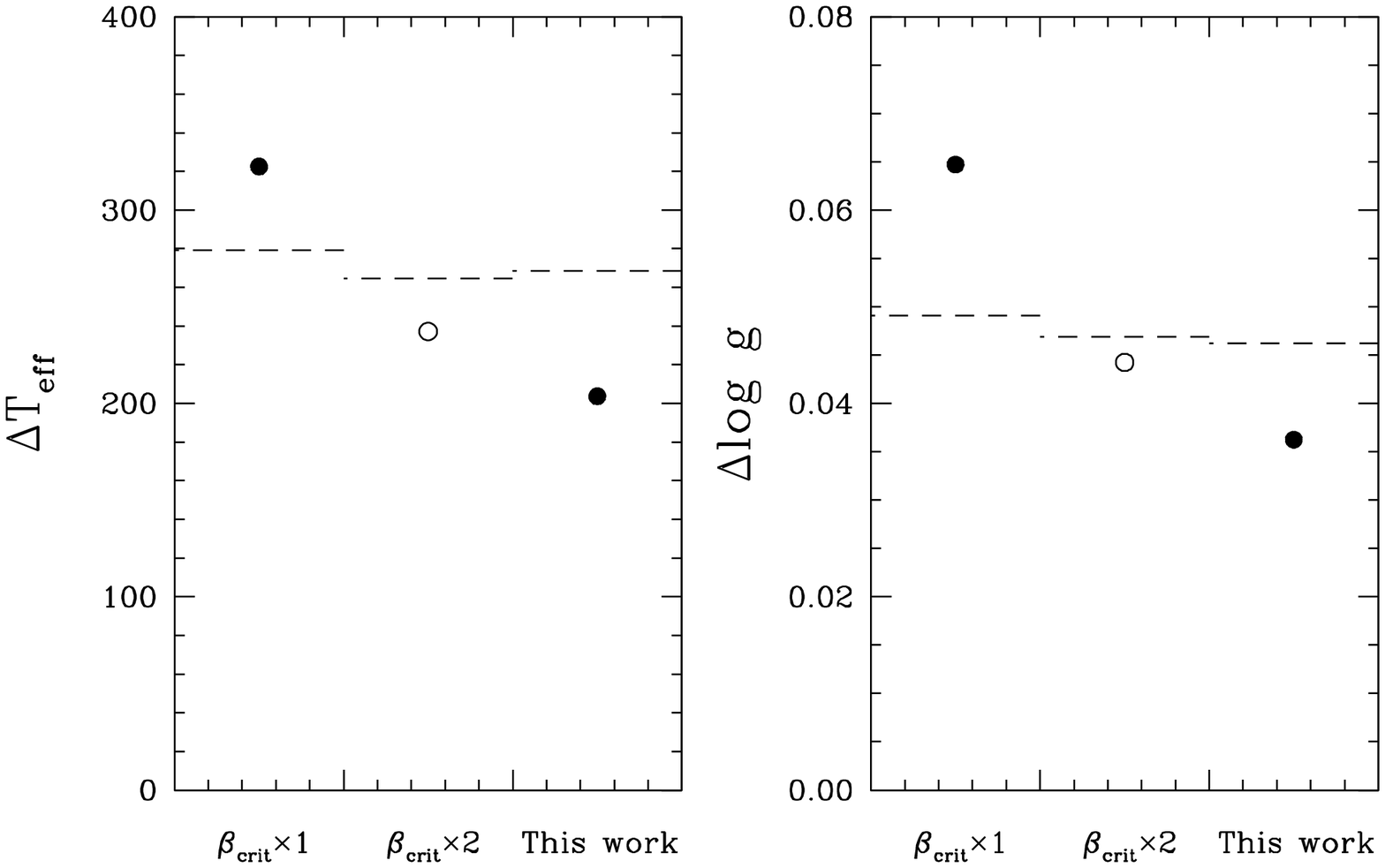]
{{\it Left:} For each star in the PG sample with 40,000~K $> T_{\rm
eff} > 13,000$~K, we computed the average absolute deviation in
$\Te$ between solutions obtained from fits that include 2 to 5
lines (similar to Fig.~\ref{fg:f2}). These deviations were then averaged
for all DA stars in our sample ({\it filled dots}). The results
are shown for the three grids discussed in the text and identified on
the $x-$axis. For comparison, the mean uncertainty of the fitting
procedure is shown as dotted lines. {\it Right:} Same as left panel
but for the dispersion in $\log$ g.
\label{fg:f9}}

\figcaption[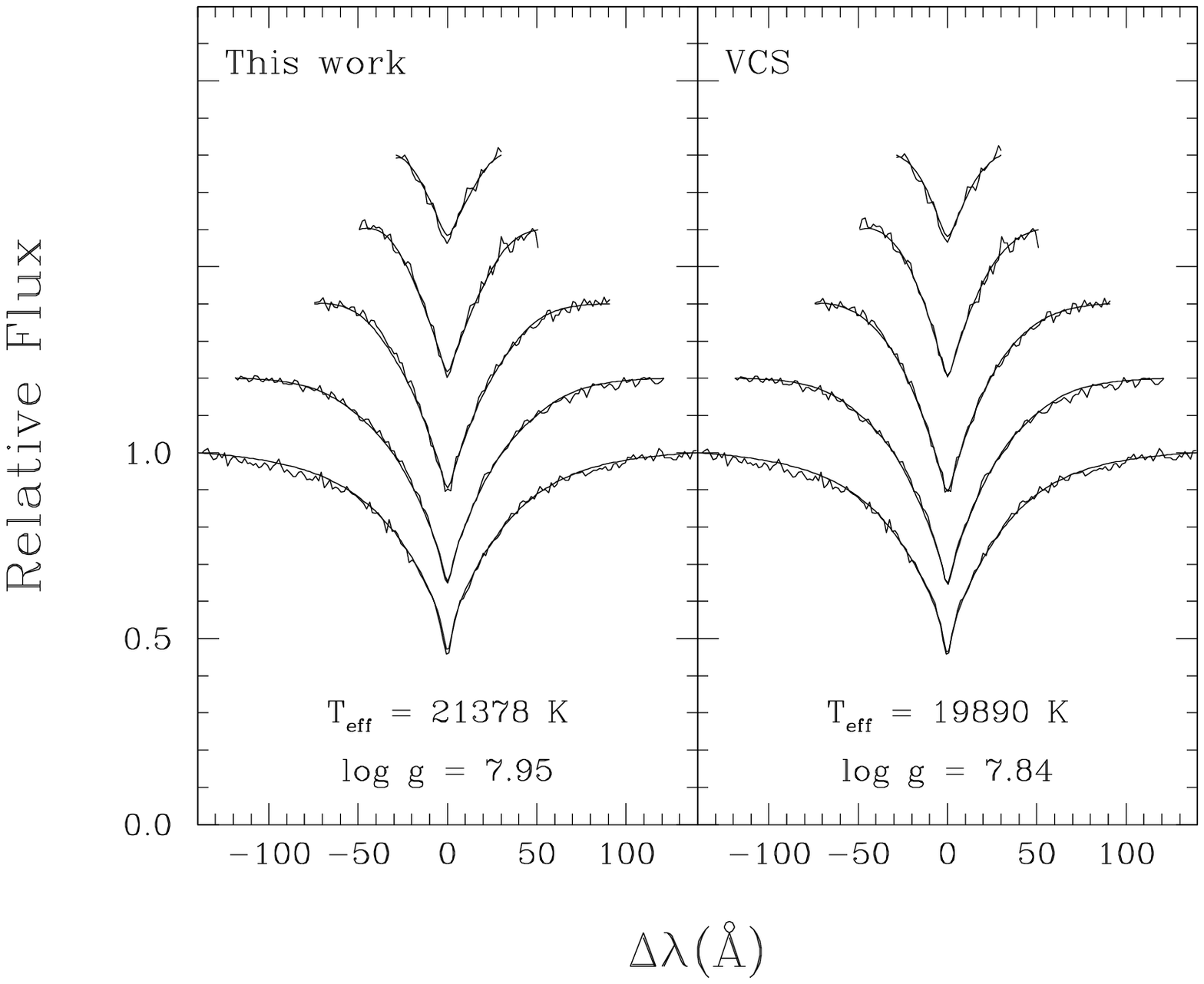] 
{Our best fit to the Balmer lines of WD 0205+250 with our improved
line profiles ({\it left panel}) and with the VCS
profiles ({\it right panel}). The atmospheric parameters are given in
each panel.
\label{fg:f10}}

\figcaption[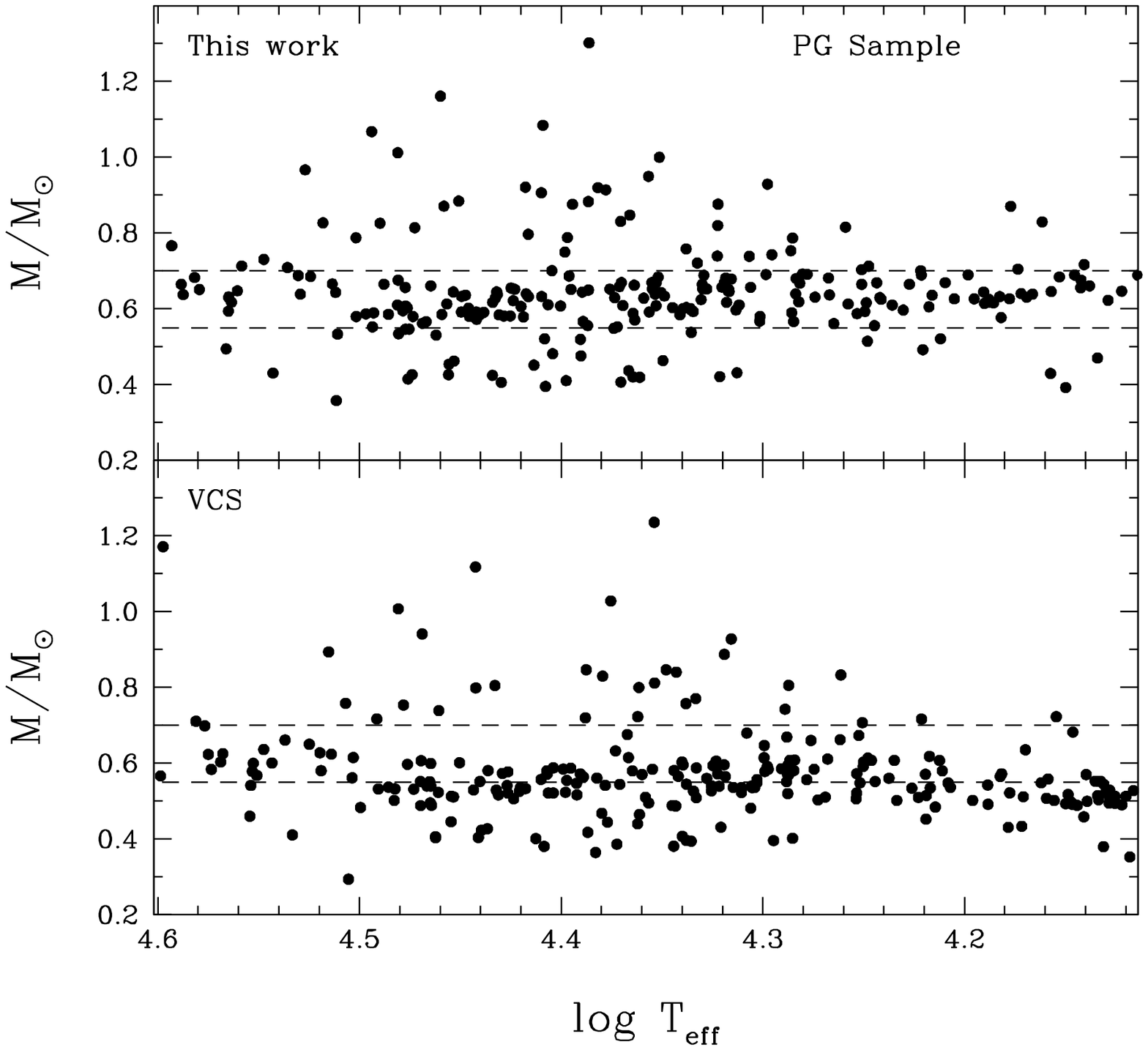] 
{Mass versus $\Te$ distribution for the DA stars in the PG
sample in the range 40,000 K $> \Te >$ 13,000 K. Results are
shown for both our improved lines profiles ({\it top panel}) and the
VCS profiles ({\it bottom panel}). Lines of constant mass at 0.55 and
0.70 \msun\ are shown as a reference.
\label{fg:f11}}

\figcaption[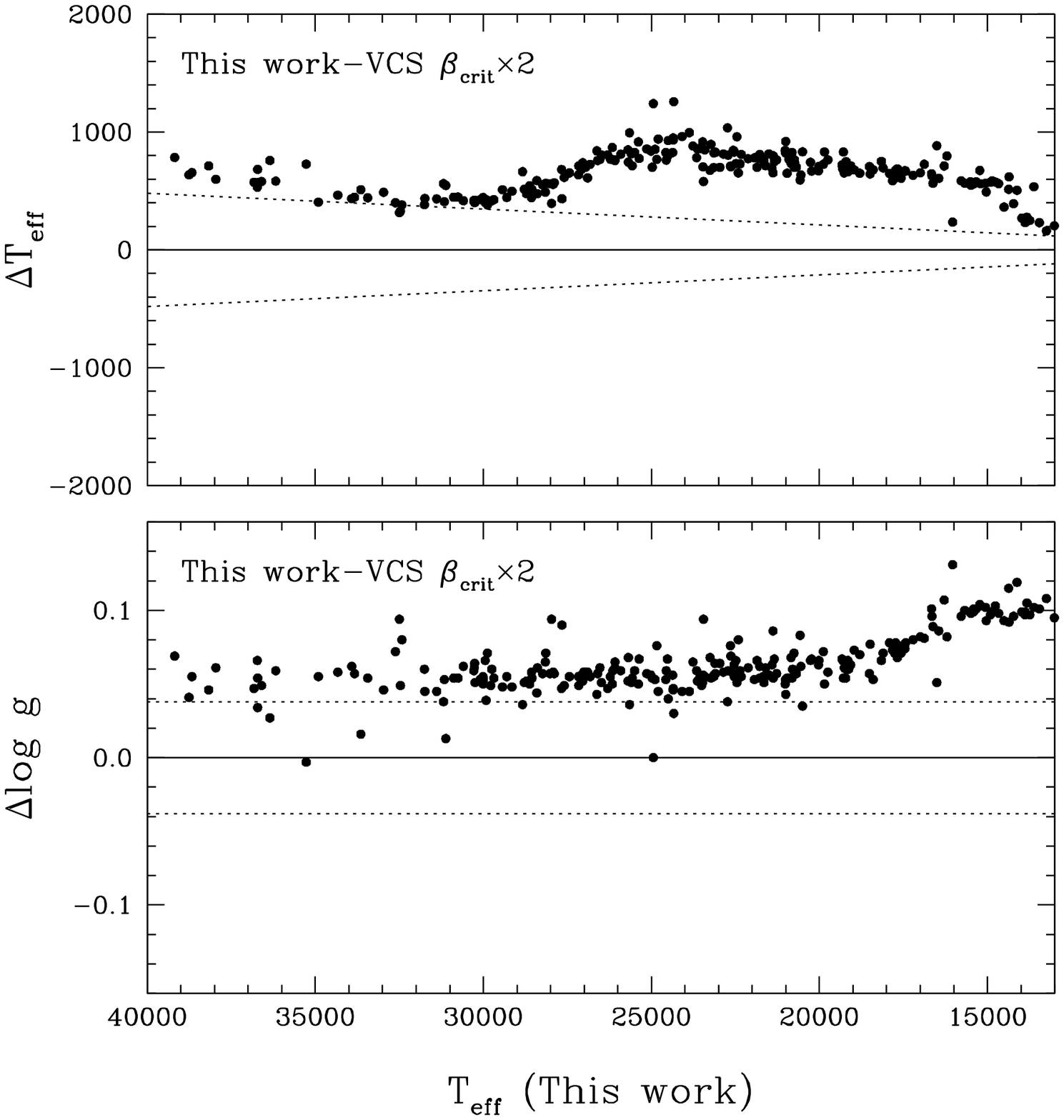] 
{Differences in $\Te$ and $\logg$ obtained with our improved models
and with the VCS profiles ($\beta_{\rm crit} \times 2$) as a function
of effective temperature for the DA stars in the PG sample in the
range 40,000 K $> \Te >$ 13,000 K. The solid lines represent the 1:1
correlation while the dotted lines correspond to the uncertainties
of the spectroscopic method as determined by \citet{liebert05}.
\label{fg:f12}}

\figcaption[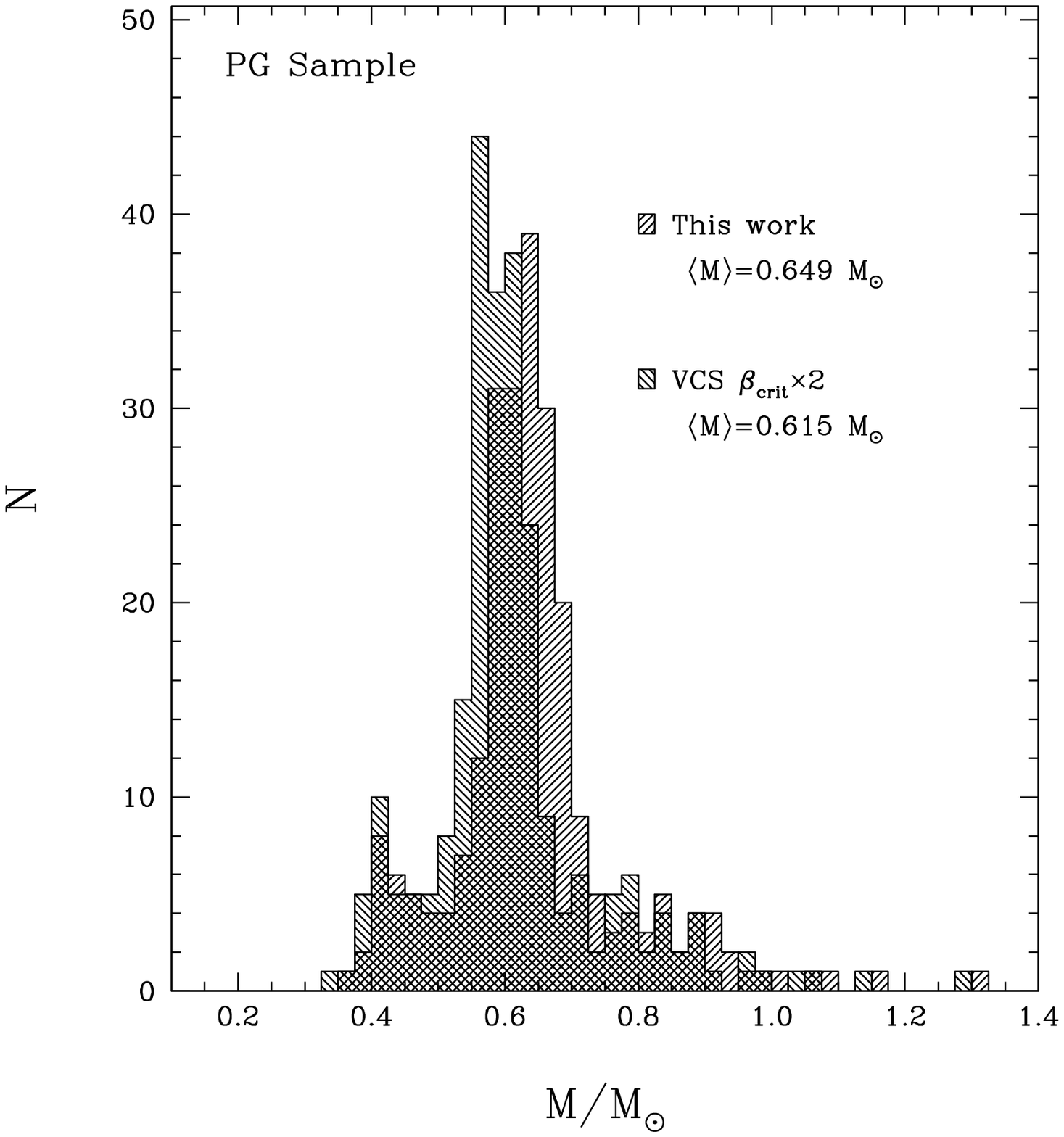] 
{Mass distributions for the subsample of PG stars studied in Figure
\ref{fg:f12}. The mean masses are reported in the figure.
\label{fg:f13}}

\figcaption[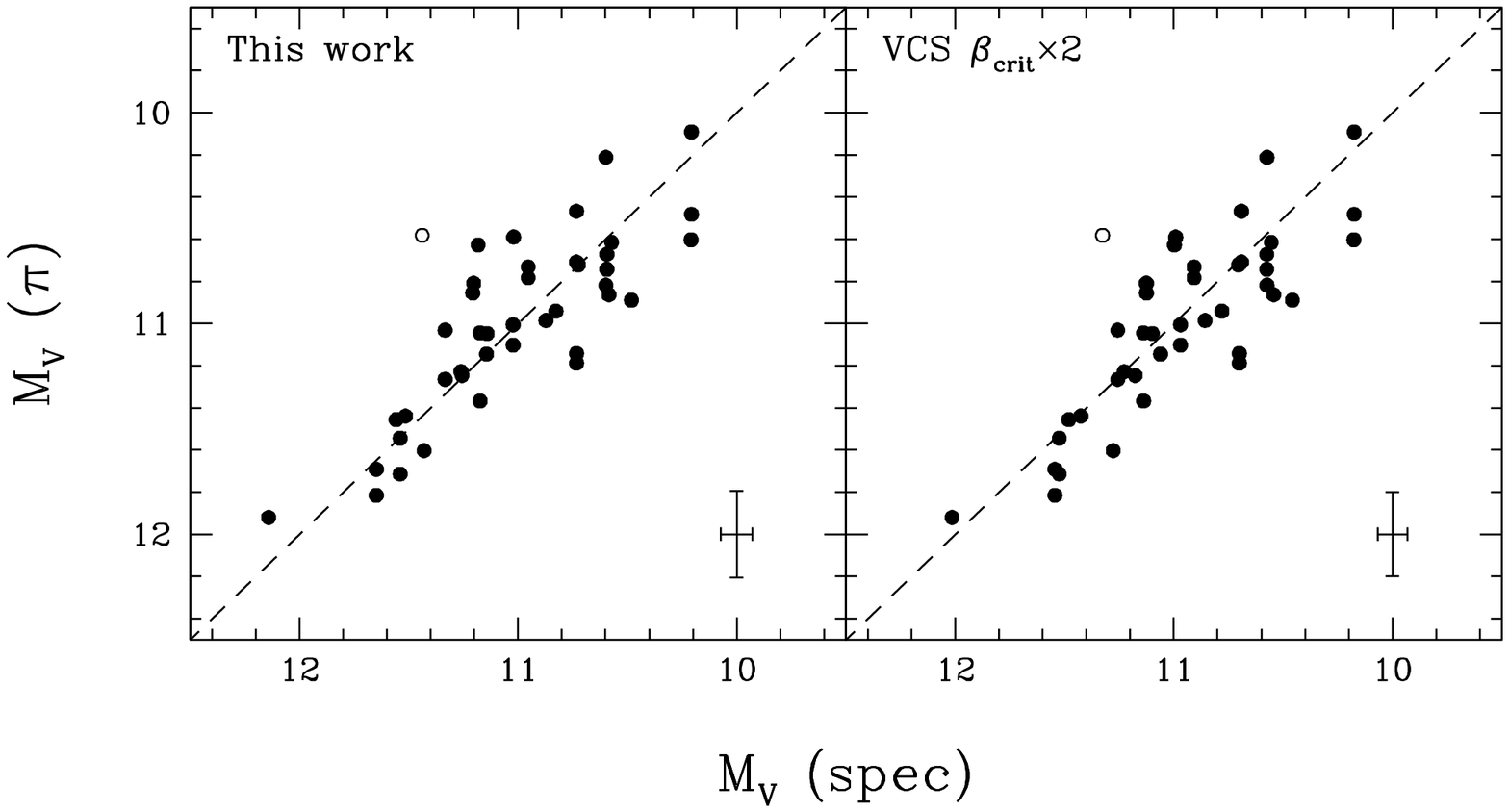] 
{Comparison of the absolute visual magnitudes obtained from
trigonometric parallax measurements and from the spectroscopic
technique using both model grids. The analysis is restricted to 40,000
K $> \Te >$ 13,000 K. The error bars represent the average
uncertainties. WD~1606+422 ({\it open circle}) is a suspected double
degenerate binary \citep{bergeron01}. \label{fg:f14}}

\begin{figure}[p]
\plotone{f1.eps}
\begin{flushright}
Figure \ref{fg:f1}
\end{flushright}
\end{figure}

\begin{figure}[p]
\plotone{f2.eps}
\begin{flushright}
Figure \ref{fg:f2}
\end{flushright}
\end{figure}

\begin{figure}[p]
\plotone{f3.eps}
\begin{flushright}
Figure \ref{fg:f3}
\end{flushright}
\end{figure}

\begin{figure}[p]
\plotone{f4.eps}
\begin{flushright}
Figure \ref{fg:f4}
\end{flushright}
\end{figure}

\begin{figure}[p]
\plotone{f5.eps}
\begin{flushright}
Figure \ref{fg:f5}
\end{flushright}
\end{figure}

\begin{figure}[p]
\plotone{f6.eps}
\begin{flushright}
Figure \ref{fg:f6}
\end{flushright}
\end{figure}

\begin{figure}[p]
\plotone{f7.eps}
\begin{flushright}
Figure \ref{fg:f7}
\end{flushright}
\end{figure}

\begin{figure}[p]
\plotone{f8.eps}
\begin{flushright}
Figure \ref{fg:f8}
\end{flushright}
\end{figure}

\begin{figure}[p]
\plotone{f9.eps}
\begin{flushright}
Figure \ref{fg:f9}
\end{flushright}
\end{figure}

\begin{figure}[p]
\plotone{f10.eps}
\begin{flushright}
Figure \ref{fg:f10}
\end{flushright}
\end{figure}

\begin{figure}[p]
\plotone{f11.eps}
\begin{flushright}
Figure \ref{fg:f11}
\end{flushright}
\end{figure}

\begin{figure}[p]
\plotone{f12.eps}
\begin{flushright}
Figure \ref{fg:f12}
\end{flushright}
\end{figure}

\begin{figure}[p]
\plotone{f13.eps}
\begin{flushright}
Figure \ref{fg:f13}
\end{flushright}
\end{figure}

\begin{figure}[p]
\plotone{f14.eps}
\begin{flushright}
Figure \ref{fg:f14}
\end{flushright}
\end{figure}

\end{document}

%% file: tab1.tex
\clearpage
\begin{deluxetable}{lcrrc}
\tabletypesize{\scriptsize}
\tablecolumns{5}
\tablewidth{0pt}
\tablecaption{Opacity Sources Included}
\tablehead{
\colhead{Opacity} &
\colhead{} &
\colhead{Populations} &
\colhead{Cross Section} 
}
\startdata
H bound-bound, Stark broadening (H--H$^{+}$, H$-e^{-}$)  & & N$_{i}($H$)$ & This work \\
H bound-bound, quasi-molecular (H--H, H--H$^{+}$)   & & N(H) & \citet{allard04}\\
H bound-bound, neutral broadening (H--H)  & & N$_{i}($H$)$ & \citet{ali65,ali66} \\
H bound-free   & & N$_{i}$(H) & \citet{mihalas78} \\
H free-free   & & N$_e$N(H$^+$) & \citet{mihalas78} \\
H$_2$ free-free   & & N$_e$N(H$_2^+$) & we assume that of H I \\
H$_3$ free-free   & & N$_e$N(H$_3^+$) & we assume that of H I \\
H$^{-}$ bound-free & & N(H$^{-}$) & \citet{john88}\\
H$^{-}$ free-free & & N$_e$N(H) & \citet{john88}\\
H$_2 ^{+}$ bound-free & & N(H$_2 ^{+}$) & \citet{kurucz70}\\
H$_2 ^{+}$ free-free & & N(H)N(H$^+$) & \citet{kurucz70}\\
H$_2 ^{-}$ free-free & & N$_e$N(H$_2$) & \citet{bell80}\\
CIA H-H$_2$ & & N(H)N(H$_2$) & \citet{gustafsson03} \\
CIA H$_2$-H$_2$ & & N(H$_2$)N(H$_2$) & \citet{borysow01}\\
Rayleigh H & & N(H)& \citet{kissel00}\\
Rayleigh H$_2$ & & N(H$_2$)& \citet{dalgarno62}\\
Thompson $e$ scattering & & N$_e$& \citet{mihalas78}\\
\enddata
\tablecomments{The bound-bound H$_2$ and H$_3^{+}$ opacities are always
negligible in DA white dwarfs.}
\end{deluxetable}